# *Mémoire de Fin d'Etudes*

*Pour l'Obtention du Diplôme de*
*Master en Informatique*

**Présenté par :**
**SADJI SAFIA**

*Domaine : Mathématiques & Informatique*
*Spécialité : AD-SI*

**Session Juin 2018**

**THEME**

UN PROTOCOLE DE NEGOCIATION POUR LES SYSTEMES D'AIDE A LA DECISION DE GROUPE.

Encadré par : Mme HAMDADOU D
Co-encadré par : Mr KHIAT S

**Jury**

Président : Mr ZEKRI
Examinateur : Mr ABDI

**CodeMaster : 02/2018**

*Promotion 2017/2018*

# *Remerciements*

*Tout d'abord et avant tout, je remercie le bon DIEU de m'avoir donné la force et le courage pour réaliser ce modeste travail, de m'avoir illuminé et guidé sur le chemin du bien.*

*Je tiens à remercier mon encadrante Mme HAMDADOU Djamila, pour l'orientation, la confiance, et la patience qu'elle m'a donné.*

*Je remercie mon Co-encadrant Mr KHIAT Sofiane, pour ses précieux conseils, et sa présence tout au long de la réalisation de ce projet de fin d'étude.*

*Je remercie chaleureusement Mr ZEKRI Lougmiri, pour ses conseils durant ces deux années, et qui m'a honoré en acceptant d'être le présidant de mon jury de soutenance.*

*Je voudrais aussi exprimer ma gratitude à Mr ABDI Mustapha Kamel, qui a bien voulu prendre de son temps pour évaluer mon travail et donner son avis.*

*Merci à Melle OUFELLA Sarah pour ses aides et orientations.*

*Et un grand merci à ma sœur BENHADDOUCHE Douaa, pour ses conseils, et sa présence quand j'en avais besoin*

*Je voudrais, également, remercier mon enseignante Mme MECHACHE Kheira pour son aide au bon moment*

# إهداء

**الحمد لله الذي تتم بفضله الصالحات،والحمد لله كما ينبغي لجلال وجهه وعظيم سلطانه.**

إلى المرأة التي فدت حياتها لحمايتي،

إلى المرأة التي جاهدت لتربيتي، وسهرت لتعليمي،

إلى أمي حبيبتي، جنتي في الأرض، اسأل الله لكي أن يحفظك، وأن يسترك، ويرزقك طول العمر وحسن العمل، فجزاك الله عنا بجنتا الدنيا والآخرة.

إلى أبي رحمه الله،

إلى كل إخوتي وأخواتي،

إلى كل من ساعدني يوما ولو بكلمة طيبة جزآكم الله كل خير عني.

ساجي صافية

## Liste des Figures

# Sommaire

## Chapitre II : Négociation dans les SMA

## Liste des Abréviations

| | |
|---:|---|
| *ACL* | **A**gent **C**ommunication **L**anguages |
| *AHP* | **A**nalycal **H**ierarchy **P**rocess |
| *DSS* | **D**ecision **S**upport **S**ystem |
| *FIPA* | **F**oundation for **I**ntelligent **P**hysical **A**gents |
| *PROMETHEE* | **P**reference **R**anking **O**rganisation **METH**od for **E**nrichment **E**valuations |
| *MP* | **M**atrice de **P**erformance |
| *GDSS* | **G**roup **D**ecision **S**upport **S**ystem |
| *SMA* | **S**ystème **M**ulti-**A**gents |
| *SIAD* | **S**ystème **I**nteractif d'**A**ide à la **D**écision |
| *KQML* | **K**nowledge **Q**uery **M**anipulation **L**anguage |

**Résumé.**

Notre contribution concerne les systèmes interactifs d'aide à la décision multi participants pour la prise de la décision de groupe. Nous nous appliquons, à travers cette étude, à mettre en place une démarche décisionnelle visant à représenter la multiplicité des décideurs, leurs diversités, leurs comportements ainsi que leurs interactions.

Dans cette optique, nous contribuons à la conception et l'élaboration d'un système d'aide à la décision de groupe. Ce dernier est modélisé par un système multi agents tout en exploitant un protocole de négociation basé sur la médiation et la concession. . Ce protocole permet aux décideurs d'exprimer leurs préférences en utilisant les méthodes d'analyse multicritères, principalement la méthode d'agrégation totale AHP (Analyse Hiérarchique des Procédés) et la méthode d'agrégation partielle PROMETHEE II.

**Mots clés.**

Aide à la décision de groupe, Analyse multicritères, AHP, PROMETHEE, Système Multi-Agents, Protocole de négociation, Concession Monotone.

**Abstract.**

Our contribution concerns interactive decision support systems for group decision support. Through this study, we apply to implement a decisional process aiming to represent the multiplicity of actors, their diversity, their behaviors and their interactions.

In this context, we contribute to the design and development of a group decision support system. The latter is modeled by a multi agents system while exploiting a negotiation protocol based on mediation and concession. This protocol allows decision-makers to express their preferences using multicriteria analysis methods, mainly the method by total aggregation AHP (Hierarchical Process Analysis) and the method by partial aggregation PROMETHEE II .

# Introduction générale

**Contexte et Problématique**

La décision constitue une grande partie des activités des humaines, elle est généralement basée sur la qualité des informations, les connaissances et les expériences des décideurs. Cette décision peut être un choix parmi les solutions possibles (alternatives) proposées, ou une nouvelle solution, et cela revient au type du problème décisionnel étudié.

L'aide à la décision de groupe vise à résoudre des problèmes très confus, elle consiste à créer un système ou un outil informatique appelé un système d'aide à la décision SIAD de groupe, en Anglais GDSS (Group Decision Support System) implémenté pour assister les décideurs et les aider à mieux exprimer leurs préférences, et subjectivités, vis-à-vis un problème donné en facilitant la prise de décision par proposition des éléments de réponses.

Dans cette optique, la notion de négociation, constitue, actuellement, le sujet d'un très grand nombre de travaux de recherche. Elle est définit comme étant un processus dont l'objectif est de mettre plusieurs entités (personnes, décideurs, agents, etc.) en discussion afin d'arriver à un accord commun. Elle prend plusieurs formes, et le choix entre ces formes dépend, principalement, du problème décisionnel abordé.

A cet effet, les décideurs font souvent face à des problématiques décisionnelles multicritères impliquant des critères souvent contradictoires. Ce type de problématiques est réputé d'être complexe, faisant intervenir plusieurs acteurs (plusieurs personnes et institutions), ayant des préférences et des points de vus différentes.

La problématique abordée dans la présente étude peut être formulée selon les questions suivantes :

- *Comment représenter la multiplicité et la, diversité des décideurs dans un système d'aide à la décision de groupe?,*
- *Comment représenter la multiplicité et la diversité des critères dans un GDSS?,*
- *Quelle stratégie de négociation faut-il adopter entre les décideurs?*

## Contribution

La résolution des problèmes décisionnels impliquant plusieurs décideurs, et plusieurs critères contradictoires, consiste à trouver une décision commune mutuellement acceptable. D'où la

# Introduction Générale

nécessité d'un processus de négociation permettant de trouver un accord commun pour un groupe d'agents, face à un conflit sur la décision à prendre.

Notre objectif, à travers ce projet de fin d'études, est de mettre en place un outil qui permet d'aider un ensemble de décideurs, à mieux exprimer leurs préférences, et à trouver une solution acceptable par tous les décideurs en mettant en œuvre un processus de négociation .

Un autre objectif visé par la présente étude est la proposition d'un nouveau protocole de négociation qui sera intégré dans un système multi-agents modélisant le système d'aide à la décision de groupe. Le protocole proposé est basé sur la concession monotone et exploite les avantages de l'analyse multicritères en utilisant, principalement, deux méthodes, l'une procédant par agrégation totale (à savoir AHP) et une autre procédant par agrégation partielle (à savoir PROMETHEE II).

## Organisation du mémoire

Ce mémoire est organisé en quatre chapitres comme suit :

### Chapitre I : L'Aide à la décision de groupe

Dans ce chapitre, nous présentons les différents concepts et définitions se rapportant à l'aide à la décision en général, et l'aide à la décision de groupe en particulier.

Au début, nous donnons un aperçu sur les concepts fondamentaux de l'aide à la décision, le processus d'aide à la décision ainsi que les acteurs impliqués, Ensuite, nous présentons, en détails, les concepts de l'aide à la décision de groupe.

### Chapitre II: La négociation dans les Système Multi-agents

Dans ce chapitre, nous présentons les systèmes multi-agents, et leurs concepts de base, Ensuite nous présentons les différents types d'interactions dans un SMA, puis, le concept de la négociation dans un SMA est détaillé.

### Chapitre III: Conception et modélisation

Ce chapitre aborde la méthodologie de conception et de modélisation de notre proposition. Une attention particulière est réservée à l'architecture globale du système ainsi que la démarche décisionnelle adoptée. Nous illustrons, également, les différents diagrammes UML associés à notre contribution.

**Chapitre IV : Mise en œuvre**

Ce chapitre est consacré à la présentation du système d'aide à la décision de groupe proposé ainsi que l'environnement de son développement. Un scénario de problème décisionnel de groupe est présenté et discuté dans ce chapitre.

**Annexe : Analyse multicritères**

Cette annexe est dédiée à la présentation de l'analyse multicritères ainsi que les principales méthodes multicritères connues dans la littérature.

Ce mémoire s'achève par une conclusion résumant les principaux apports de cette étude et ouvre des directions aux travaux futurs envisagés.

# Chapitre I

# L'Aide à la Décision de Groupe

## 1. Introduction

Prendre une décision dans notre vie quotidienne peut être quelquefois rapide ou facile pour n'importe quel corps ou n'importe qui fait face à des problèmes qu'il doit résoudre, il peut utiliser son intelligence, expérience, ou par des cas déjà passés.

Toutefois, il y a des problèmes très confus, où prendre une décision est très difficile et compliqué, et cela est dû à plusieurs facteurs [NAC, 14] :

- Une complexité structurelle des décisions.
- Un grand nombre d'alternatives dû à la complexité du problème.
- L'impact de la décision prise peut être très important, il peut être d'ordre économique, politique, organisationnel, environnement, etc.
- La nécessité de la rapidité dans la prise de la décision, c'est le cas d'urgence médicales, ou militaires, ou encore le diagnostic des installations logicielles.

Ce chapitre présente une introduction des concepts et des termes se rapportant à la décision, l'aide à la décision, et l'aide à la décision de groupe.

## 2. Décision

Pour mieux cerner la notion de décision, nous allons présenter ici des définitions proposées par des auteurs différents :

*« Une décision est une action qui est prise pour faire face à une difficulté ou répondre à une modification de l'environnement, c'est-à-dire, pour résoudre une problème qui se pose à l'individu ou à l'organisation »* [ADL, 10].

Selon HAMDADOU [HAM, 08] *« La décision est un choix entre les actions, les solutions ainsi que la ou les alternatives »*.

Et pour la plupart, *« Une décision c'est le résultat d'un processus mental qui choisit une parmi plusieurs alternatives, mutuellement exclusives »* [BOU, 12].

Et pour les autres, *« la décision concerne aussi le processus de sélection de buts et d'alternatives »*.

## 2.1. Les étapes de la décision

La décision est un processus de résolution de problème qui met en œuvre des connaissances de natures très variées, la décision simple correspond à la classe de « problèmes relativement bien défini » [MAD, 11]. Un problème existe si [MAD, 11] :

- Le décideur perçoit un état interne ou externe non désiré A ;
- L'état de départ A non désiré doit être transformé en état de but B ;
- Le décideur ne sait pas au départ comment parcourir le chemin qui va du point de départ A vers le point de solution B.

## 2.2. Approches selon la nature des variables de décision

Les décisions sont de deux natures : les décisions programmables ou les décisions non programmables [GBA, 12]:

- **Les décisions programmables**: ce sont des décisions faciles à prendre qui portent sur des variables quantitatives et peu nombreuses, car il est facile de formaliser la décision par l'élaboration d'un algorithme.
- **Les décisions non programmables** : ce sont des décisions difficiles à prendre pour lesquelles les variables sont qualitatives et nombreuses. Il est difficile de les inclure dans un modèle mathématique.

## 2.3. Les modèles de décision

Les modèles de décision reposent généralement sur des hypothèses de rationalité de décideur et les alternatives, on va présenter deux types de modèles les modèles normatifs, et les modèles descriptifs [NAC, 14].

### 2.3.1. Les modèles normatifs

Ils fournissent des solutions optimales. Dans ce type de modèle, l'espace de recherche va être tout consulté et exploré, on peut citer trois catégories de modèles normatifs [NAC, 14] :

1. **Enumération complète** : ces modèles cherchent des meilleures solutions parmi un ensemble relativement petit des alternatives. Les principales méthodes sont des tables, les arbres de décision, etc.
2. **Optimisation par des algorithmes** : il s'agit de trouver la meilleure solution parmi un ensemble important voire même infini d'alternatives, en utilisant un processus d'amélioration pas à pas. Les principales méthodes sont la programmation linéaire, programmation linéaire en nombre entier, etc.
3. **Optimisation via des formules analytiques** : c'est une optimisation qui recherche la meilleure solution en une seule étape, en utilisant une formule analytique.

### 2.3.2. Les modèles descriptifs

Ils donnent une solution satisfaisante en explorant une partie des solutions, Parmi les modèles descriptifs, nous citons trois catégorie [NAC, 14]:

1. **La simulation :** c'est une technique qui permet de mener des expériences de prise la décision en observant les caractéristiques d'un système donné sous déférentes configurations. Cette technique permet d'aboutir à une solution en choisissant la meilleure parmi les alternatives évaluées.
2. **La prédiction** : cette technique permet de prévoir les conséquences des différentes alternatives selon des modèles de prédiction. Les modèles markoviens font partie des méthodes les plus connues de cette catégorie, La prédiction fournit une assez bonne solution ou une solution satisfaisante.
3. **Les heuristiques** : elles permettent d'atteindre une solution satisfaisante à moindre cout en utilisant les techniques de la programmation heuristique et les systèmes à base de connaissances. Ces méthodes sont plutôt utilisées pour des problèmes complexes et mal structurés ou la détermination de solutions peut entrainer un cout et un temps élevé.

## 3. L'Aide à la décision

L'aide à la décision est définie selon B.ROY [BOU, 12] comme étant:

*« L'activité de celui qui, prenant appui sur des modèles clairement explicités mais non nécessairement complètement formalisés, aide à obtenir des éléments de réponse aux questions que se pose un intervenant dans un processus de décision, éléments concourant à éclairer la décision et normalement à recommander , ou simplement à favoriser, un comportement de nature à accroître la cohérence entre l'évolution du processus d'une part, les objectifs et le système de valeurs au service desquels cet intervenant se trouve placé d'autre part ».*

L'aide à la décision utilise des techniques et des méthodologies issues du domaine des mathématiques appliquées telles que l'optimisation, les statistiques, la théorie de la décision ainsi que des théories de domaines moins formels telles que l'analyse des organisations et les sciences cognitives [PAS, 05].

D'après l'auteur dans [HAM, 08] *« l'aide à la décision correspond à une démarche constructive dans laquelle on considère que les préférences des intervenants sont souvent conflictuelle et peu structurées, appelées à évolué au sein d'un processus et influencées de fait même de la mise en œuvre du modèle. Le système d'aide à la décision est alors élaboré en cherchant à tirer la partie de la perception de problèmes. »*

### 3.1. Les acteurs d'aide à la décision

Un acteur en aide à la décision est tous individus intervenant dans le processus d'aide à la décision. HAMDADOU [HAM, 08], [HAM, 16], distingue plusieurs acteurs dont nous citons :

- **Le décideur:** la personne (ou les personnes) assistée(s) par l'aide à l'AD et qui est aidée pour mieux exprimer ses préférences vis-à-vis une situation donnée.
- **L'homme d'étude (l'analyste) :** un individu ou un groupe d'individus qui a pour rôle d'établir le système de préférence, de définir le modèle d'aide à la décision de l'exploiter afin d'obtenir des réponses et d'établir des recommandations pour conseiller le décideur sur les solutions envisageables, l'homme d'étude est à distinguer du négociateur et de médiateur.
- **Les intervenants:** sont ceux qui, de par leur intervention, conditionnent directement la décision en fonction du système de valeurs dont ils sont porteurs.
- **Les Agis:** ils sont concernés par les conséquences de la décision, ils interviennent directement dans le processus par l'image que d'autres acteurs se font de leur valeurs et plus concrètement de leurs systèmes de préférences.
- **Les demandeurs:** ils demandent l'étude et allouent les moyens.

- **Le négociateur:** mandaté par un décideur en vue de faire valoir la position de celui-ci dans une négociation et de rechercher une action compromis.
- **Le médiateur:** intervient en vue d'aider le décideur(ou les négociateurs) à rechercher le compromis [HAM, 16].
- **L'arbitre (le juge):** intervient en se substituant aux acteurs dans la recherche du compromis.

## 4. Processus d'aide à la décision

Un processus d'aide à la décision est l'ensemble des étapes et des activités qui commence par un problème, pouvant être posé comme une question dans la majorité des cas, et se termine par un ensemble d'alternatives qui peuvent donner une solution dont le décideur a besoin. SIMON a proposé le modèle canonique suivant (Figure I.1) procédant en quatre étapes [PAS, 05]:

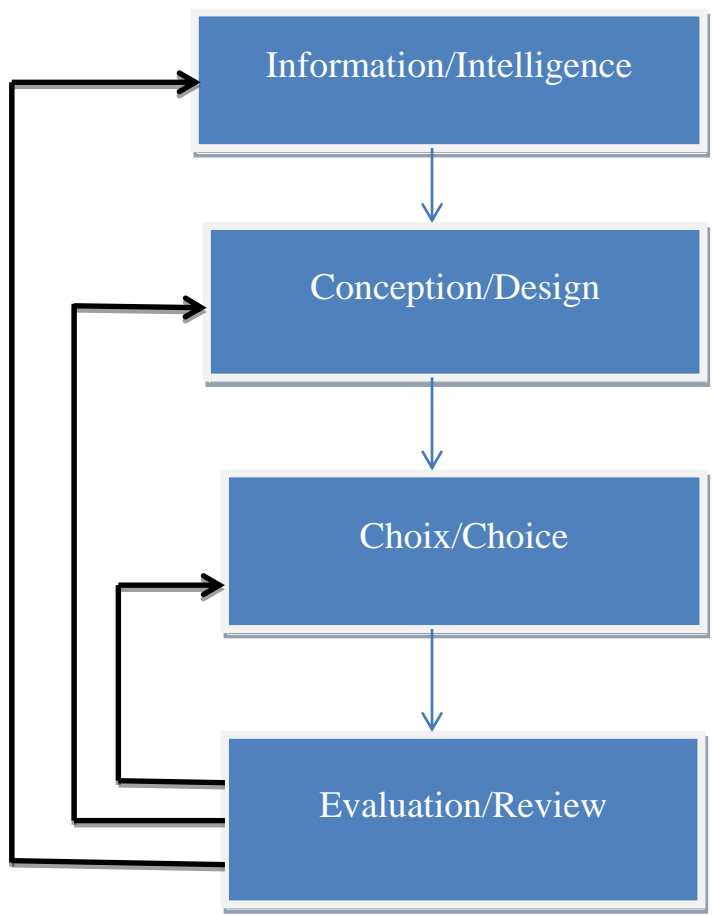

**Figure I.1** : Le modèle de processus de décisionproposé par SIMON.

Ces étapes sont définies comme suit [NAC, 14]:

1. **L'étape d'information(Intelligence)** : c'est une phase d'identification du problème à résoudre, pour cela il est nécessaire de rechercher les informations pertinentes en fonction des préoccupations du décideur.
2. **L'étape de conception(Design)** : c'est une phase de modélisation proprement dite, le décideur construit des solutions et imagine des scénarios, Cette phase aboutit aux différents chemins possibles à la résolution du problème.
3. **Choix d'un mode d'action(Choice)** : c'est phase de sélection d'un mode d'action particulier, c'est-à-dire la prise d'une décision.
4. **Control ou Evaluation(Review)** : cette phase permet d'évaluer la solution choisie (la décision prise). Elle peut amener à un retour arrière vers l'une des trois phases précédentes ou, au contraire, à la validation de la solution.

## 5. Les systèmes d'aide à la décision

Jusqu'aux années 70, les systèmes de décision étaient associés à des méthodologies propres à la recherche opérationnelle, l'analyse des données et de calcul optimal. Le but de ces systèmes était de résoudre des problèmes par la recherche d'une solution optimale, en calculant le maxima ou le minima de fonctions mathématiques exprimant un objectif à atteindre. Grâce au progrès informatique, l'une des évolutions des systèmes d'aide à la décision a été de se rapprocher des utilisateurs pour leur promettre d'intervenir dans le processus de décision [YOU. xx].

Les SIAD ou en anglais DSS pour Decision support System, ont été d'abord introduit par une école anglosaxone, l'un des premiers auteurs à les avoir cités est Scott Morton, il a introduit la notion de Système de décision et de gestion (Management Decision System) [NAC, 14].

### 5.1 Les fonctionnalités des SIAD

Divers fonctions sont attribuées aux SIAD [NAC, 14] :

- Ils doivent apporter une aide pour les problèmes peu ou mal structurés, en connectant ensemble des jugements humains et des informations calculées.
- Ils doivent posséder une interface simple et conviviale afin d'éviter que l'usager soit perdu devant la complexité du système.
- Ils doivent fournir une aide pour différentes catégories des utilisateurs ou différents groupe des utilisateurs.
- Ils doivent supporter des processus interdépendants ou séquentiels.
- Ils doivent être adaptatifs dans le temps, le décideur doit être capable de supporter des conditions qui changent rapidement et d'adapter le SIAD pour faire face aux nouvelles situations. Un SIAD doit être suffisamment flexible pour que le décideur puisse ajouter, détruire, combiner, changer réarranger les variables de processus de décision ainsi que les différents calculs fournissant, ainsi une réponse rapide à des situations inattendues.

- Ils doivent laisser le contrôle de toutes les étapes du processus de décision au décideur pour que celui-ci puisse remettre en cause à tout moment les recommandations faites par les SIAD. Un SIAD doit aider le décideur et non pas se substituer à lui.
- Ils doivent utiliser des modèles, la modélisation permet d'expérimenter différentes stratégies sous différentes conditions. Ces expériences peuvent apporter de nouvelles vues sur le problème et un apprentissage.
- Les SIAD les plus avancés utilisent un système à base de connaissances qui apporte notamment une aide efficace et effective dans des problèmes nécessitant une expertise.
- Ils doivent permettre la recherche heuristique.
- Ils ne doivent pas disposerdes outils de type boite noire. Le fonctionnement d'un SIAD doit être fait de manière à ce que le décideur le comprenne et l'accepte.

## 5.2 Les différents types de SIAD

Les premiers SIAD sont apparus en 1968, ils n'ont été effectifs qu'à la fin des années soixante-dix ou divers outils d'aide à la décision sont devenus opérationnels. La définition et les spécifications d'un SIAD ne peuvent pas être fixées précisément car il n'est pas possible de trouver une approche générique pour l'ensemble des cas et des domaines possibles [FER, xx].

C'est pourquoi ils ont proposés le regroupement suivant [FER, xx] :

- **SIAD passif :** donne un avis qui ne vient pas empiéter l'autonomie de l'utilisateur.
- **SIAD traditionnel :** est un assistant dont la principale utilisation concerne les interactions de type « What-if ?».
- **SIAD étendu:** assure la fonction de consultant et se place au même niveau que l'utilisateur.
- **SIAD nominatif :** domine le processus de décision. L'opérateur ne remplit plus qu'un rôle passif dans la prise de décision.

La figure suivante( figure I.2) illustre une architecture d'un système d'aide à la décision, et son interaction avec l'utilisateur (décideur), tel que le décideur pose son problème comme une question et le système va proposer un ensemble d'alternatives, et après c'est le décideur qui décide.

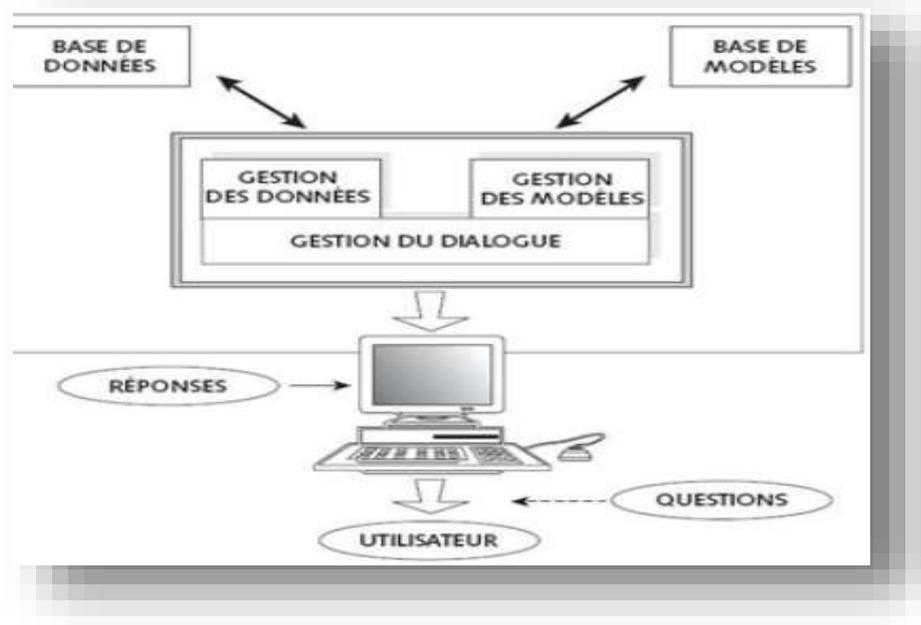

**Figure I.2** : Système d'aide à la décision.

## 6. L'aide à la décision de groupe

### 6.1. Définition

Dans la littérature, il existe plusieurs définitions, dont nous citons :

Pour Marakas dans[BOU, 12], *« l'aide à la décision collective est une activité conduite par une entité collective composée de deux ou plusieurs individus et caractérisée à la fois en terme de propriétés de l'entité collective et de celles de ses membres individuels »*.

### 6.2. Pourquoi l'aide à la décision est difficile ?

Dans de nombreuses organisations, ou des entreprises, les problèmes décisionnels ne peuvent pas être facilement faits, et ici plusieurs raisons peuvent être évoquées [BOU, 12] :

- L'évolution économique et le marché de plus en plus concurrentiel poussent sans cesse les organisations à prendre de meilleures décisions ;
- Le marché et le monde évoluent de plus en plus vite, les décisions doivent donc être prises plus rapidement ;
- La quantité d'informations et de connaissances disponibles augmente rapidement ; les organisations ont donc besoin d'outils pour les utiliser efficacement ;

- La modification de la structure hiérarchique ; les organisations ont tendance à adopter des organigrammes de plus en plus plats et à donner plus de pouvoir aux niveaux inférieurs, notamment en faisant participer les gens à des réunions ;
- La maitrise de la complexité, l'amélioration de l'efficacité, et l'évolution technologique.
- La prise en compte des spécificités géographiques (les décideurs peuvent être représentatifs de différents sites de l'organisation et avoir des expériences relatives à des contextes organisationnels, humains, et culturels différents).

### 6.3. Les SIAD de groupe (Group Decision Support System)

Un Système Interactive d'Aide à la Décision de Groupe, GDSS pour(Group Decision Support System) en anglais, est« *un ensemble des technologies qui soutiennent les activités effectuées par des décideurs organisés en un groupe* » [ABD, 12].

Un GDSS est également connu sous l'appellation de groupware, il est communément définie comme étant *« une collection de logiciels, du matériel, et de procédure, conçus pour supporter automatiquement le travail de groupe »* [MAD, 11].

Selon HAMDADOU [HAM, 16], *« un GDSS est un système interactif et informatisé qui facilite la résolution des problèmes non structurés par un ensemble de décideurs fonctionnant ensemble en tant que groupe. Il aide des groupes, particulièrement groupes des directeurs, en analysant des situations de problème et en accomplissant des tâches de prise de décision de groupe »*

D'un point de vue technologique le groupware se situe aux frontières del'informatique et des télécommunications. En effet des technologies issues de l'informatique ainsi que de la télécommunication sont utilisées[PAS, 05].

D'un point de vue système d'information, le groupware se situe aux frontières de la bureautique et de l'informatique transactionnelle. En effet, d'un point de vue utilisateur les outils de groupware font partie de la gamme des outils de bureautique et utilisent des infocentres [PAS, 05].

### 6.4. Pourquoi les SIAD de groupe

On observe, actuellement, dans les entreprises une utilisation de plus en plus fréquente des systèmes d'information pour l'aide à la décision. Cette utilisation croissante est le résultat des deux facteurs suivants :

- Tout d'abord, les micro-ordinateurs sont de plus en plus performants alors que leur coût décroît continuellement [BUI et al, 88].

- Ensuite, la sophistication ainsi que la convivialité des logiciels ont permis d'améliorer les capacités cognitives de traitement. L'utilisation de ces logiciels est devenue de plus en plus massive à tous les niveaux de l'entreprise.

Les tableurs (Lotus 1-2-3, Framework, etc.), les systèmes de gestion de bases de données, les logiciels de télécommunications ainsi que les systèmes de bureautique (traitement de texte, publication assistée par ordinateur, etc.) sont autant d'exemples illustratifs [BUI et al, 88].

Une caractéristique commune à ces logiciels réside dans le fait que les dialogues avec le décideur sont essentiellement basés sur les modes d'interaction du type "question/réponse" et de "menu à choix multiples" pour l'entrée des données, ainsi que les représentations tabulaires et graphiques pour la sortie des données. Cependant, ces logiciels ne peuvent pas assister des groupes de décideurs dans leurs relations interpersonnelles. En effet, aucun support pour les échanges d'informations ou le partage et transfert des données n'est prévu [BUI et al, 88].

### 6.5. Le groupe

Un groupe est un ensemble d'individus, de personnes, ou des objets réunis ou qui veulent se réunir, et si on dit « réunir »c'est-à-dire il ne nécessite pas toujours la présence physique du membre de ce groupe en un même lieu et en même temps, il s'agit de permettre à un groupe de communiquer sans contrainte de temps ou de lieu.

Le travail en groupe est fondé sur le groupware et s'accorde à nommer les 3C :*Communication, Collaboration, et la Coordination*, est un ensemble des (package) applications qui support plus qu'une personne travaillant dans une tache distribuée [MAD, 11].

### 6.6. Typologie des groupes

Il existe trois types de groupes [ADL, 10] :

1. **Groupe interactif :** les membres d'un groupe interactif communiquent entre eux Ets 'efforcent de poursuivre leur tâche. Dans une réunion face-à-face, seule une personne du groupe peut proposer son idée à un instant donné, car les membres du groupe ne peuvent prêter attention qu'à une seule personne à la fois. Le groupe interactif peut tirer avantage des synergies sociales, toutefois, des tractations entre les membres peuvent causer une déperdition du processus.

2. **Groupe nominal :** dans un groupe nominal, les membres du groupe travaillent séparément sur la même tâche et un des résultats est choisi comme le produit du groupe.

3. **Equipe :** un travail d'équipe combine des aspects à la fois du travail de groupe interactif et nominal. Le groupe de travail est divisé en équipes (typiquement 2-5 personnes), qui travaillent séparément. Les équipes sont de taille assez réduite pour ne pas subir les déperditions de processus des groupes larges, mais assez large pour tirer profit des synergies sociales. Cependant, les équipes peuvent subir les tractations entre membres comme dans les groupes de travail interactifs.

## 7. Les principaux acteurs dans le groupe

Généralement, le groupe des décideurs se compose de deux principaux acteurs :

### 7.1. Le facilitateur

Une composante clé dans le processus de la prise de la décision de groupe est le facilitateur (humain) de la réunion. Le facilitateur est un agent, accepté par les participants à la réunion.
Il lance et prépare les phases de processus de prise de la décision, il définit la problématique de la décision, et organise le groupe des décideurs pour le processus de prise la décision [BOU, 12].

### 7.2. Les participants

Désignant les membres du groupe des décideurs, ils sont assistés pour exprimer leurs préférences, être aidés à la prise de la décision et pour donner leurs informations concernant la problématique traitée.

## 8. Le processus d'aide à la décision de groupe

Le terme processus de prise de décision fait référence aux états successifs par lesquels passe le groupe pour arriver à la décision [ADL, 10]. Un processus de décision de groupe comprend plusieurs étapes au cours desquelles il est question :

- de cerner la problématique et de collecter des données,
- d'élaborer des règles faisant en sorte que le processus devienne un contrat entre les participants qui s'engagent ainsi à accepter le résultat qui en découle,
- d'évaluer des solutions préconisées en regard des points de vue retenus,
- de mettre en application les règles sur lesquelles les participants se sont mis d'accord et de s'enquérir de l'adéquation des recommandations.

Les observations empiriques et leur confrontation avec des études existantes permettent de soutenir que les processus de prise de décision de groupe peuvent être modélisés de façon générique illustrée par la (figure I.3) [ADL, 10].

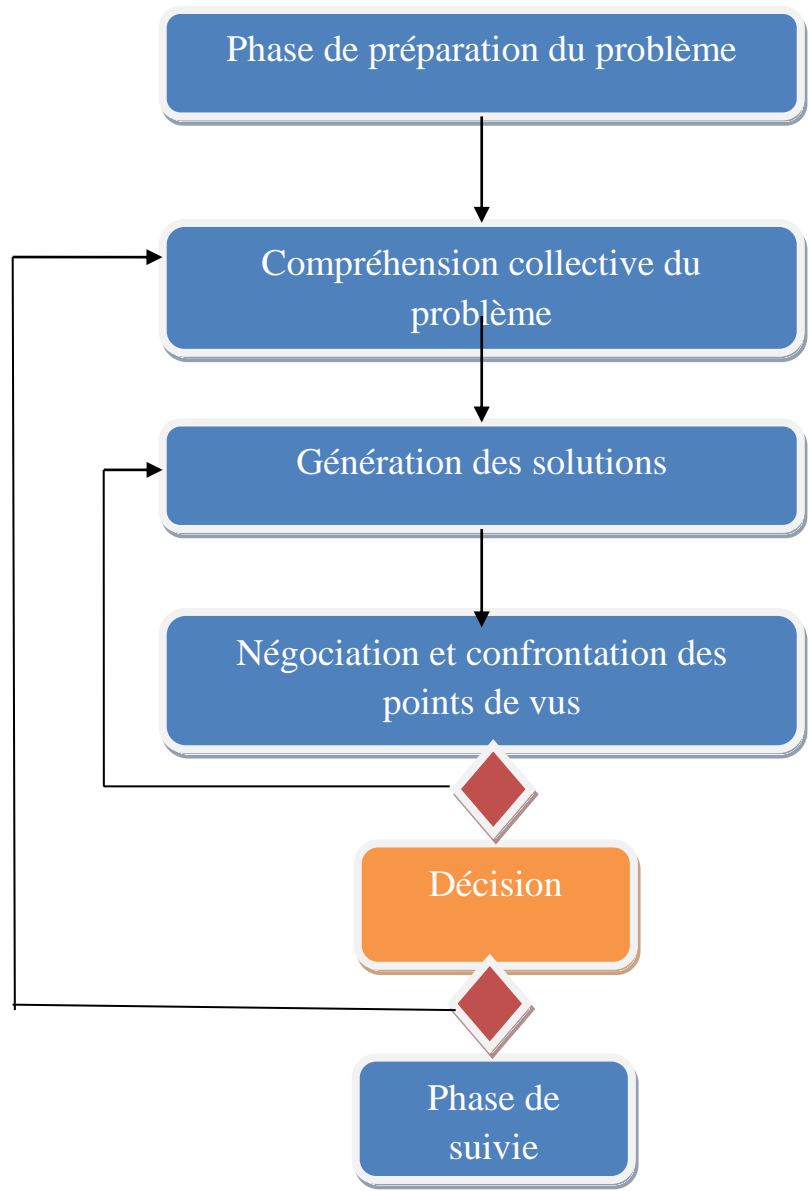

**Figure I.3 :** Modèle de processus de décision de groupe.

## 9. Les principaux avantages des GDSS

Les avantages spécifiques d'un GDSS sont nombreux, en effet, il [MAD, 11] :

- Permet la communication parallèle entre les membres des groupes.
- Offre des opportunités égales et anonymes pour contribuer par des idées et des opinions.
- Elimine les dominations trop grandes de certains membres du groupe.
- Permet de connaître les opinions communes rapidement parmi les membres du groupe.

## 10. Conclusion

La décision d'un groupe de décideurs est plus complexe que la décision d'un seul décideur, elle est considérée pour trouver et chercher un consensus entre les membres de groupe de décision.

Nous avons tenté, dans ce chapitre, de décrire le cadre général de notre travail et de connaître au mieux les concepts le structurant, et ce dans le but d'aborder et de maîtriser la problématique abordée. Nous avons, également, a donné une vue globale sur la décision, le domaine d'aide à la décision, puis, on a présenté l'aide à la décision de groupe (GDSS).

Dans le chapitre suivant, nous présentons la négociation dans les systèmes multi-agents, et ses protocoles les plus connus .

# Chapitre II

# La négociation dans les SMA

## 1. Introduction

Actuellement, la technologie d'agent est considérée comme une approche des plus importantes pour développer, modéliser, et même simuler des systèmes très complexes par des systèmes multi agents. Elle a été particulièrement reconnue comme un paradigme de la future génération des systèmes industriels, et pour tous les domaines de l'intelligence artificielle.

A ce titre, la négociation dans les SMA fait partie de nombreux travaux où il ressort un certain nombre de points communs sur lesquels il faut aujourd'hui s'appuyer.

Dans ce chapitre, nous présentons, tout d'abord, un aperçu général sur les agents, et les systèmes multi-agents(SMA) ainsi que les concepts s'y rapportant. Ensuite, nous détaillons la négociation dans les systèmes multi agents.

## 2. Notion d'agent

Plusieurs définitions présentent la notion d'agent, et il n'y a pas de définition unique pour cela, car l'agent est existant dans différents domaines telle que la Psychologie Sociale, l'Intelligence Artificielle, la science de l'Homme et de la Vie, etc.

A cet effet, plusieurs travaux ont porté sur la notion d'agent engendrant des définitions variées et riches parmi elles, on cite celle proposée par **Jacques FERBER** [OUF, 09]**:** *« Un agent est une entité réelle ou virtuelle évoluant dans un environnement capable de le percevoir et d'agir dessus, qui peut communiquer avec d'autres agents, qui exhibe un comportement autonome, lequel peut être vu comme la conséquence de ces interactions avec d'autres agents et des buts qu'il poursuit »*.

Jacques FERBER définie, également, un agent comme suit[SAH, 09] :

On appelle agent une entité physique ou virtuelle

- qui est capable d'agir dans un environnement, qui peut communiquer directement avec d'autres agents,
- qui est mue par un ensemble de tendances (sous la forme d'objectifs individuels ou d'une fonction de satisfaction, voire de survie, qu'elle cherche à optimiser)
- qui possède des ressources propres (on entend par ressource la somme des connaissances et des moyens qui sont mis à la disposition des agents afin qu'ils puissent agir dans son environnement).
- qui est capable de percevoir (mais de manière limitée) son environnement.
- qui ne dispose que d'une représentation partielle de cet environnement (et éventuellement aucune).
- qui possède des compétences et offre des services.
- qui peut éventuellement se reproduire.
- dont le comportement tend à satisfaire ses objectifs, en tenant compte des ressources et des compétences dont elle dispose, et en fonction de sa perception, de ses représentations et des communications qu'elle perçoit.

Un agent est une entité informatique, située dans un environnement, et qui agit d'une façon autonome et flexible pour atteindre les objectifs pour lesquels il a été conçu [CHA et al, 01].Les notions "situé", "autonomie" et "flexible" sont définies comme suit [CHA et al, 01]:

- *situé*: *l'agent est capable d'agir sur son environnement à partir des entrées sensorielles qu'il reçoit de ce même environnement. Exemples : systèmes de contrôle de processus, systèmes embarqués, etc.*
- *autonome*: *l'agent est capable d'agir sans l'intervention d'un tiers (humain ou agent) et contrôle ses propres actions ainsi que son état interne;*
- *flexible*: *l'agent dans ce cas est :*
  - *capable de répondre à temps : l'agent doit être capable de percevoir son environnement et élaborer une réponse dans les temps requis;*
  - *proactif : l'agent doit exhiber un comportement proactif et opportuniste, tout en étant capable de prendre l'initiative au "bon" moment;*

- *social : l'agent doit être capable d'interagir avec les autres agents (logiciels et humains) quand la situation l'exige afin de compléter ses tâches ou aider ces agents à accomplir les leurs.*

### 2.1. Les architectures d'agent

On cite trois importantes architectures pour les agents logiciels [OUF, 09]:

### 2.1.1. Architecture réactive

Les agents réactifs sont les plus faciles à concevoir et peuvent être capables d'action de groupe complexe et coordonné. Ce type d'agent n'a pas de représentation, son comportement est décrit par des règles simples de type stimulus/réponse.

### 2.1.2. Architecture cognitive

C'est un type d'agent plus complexe, doté de capacités de raisonnement importantes et disposant d'une représentation de son environnement. Chaque agent est muni d'une base de connaissances comprenant un ensemble d'informations et des savoir-faire nécessaires à la réalisation de sa tâche ainsi qu'à la gestion des interactions avec les autres agents et avec son environnement.

### 2.1.3. Architecture hybride

L'architecture hybride est en quelque sorte un compromis entre un agent purement réactif et un agent purement cognitif. Cette vision permet de concilier les capacités intéressantes des deux types d'agents précédents.

Il existe d'autres architectures des agents telles que l'architecture BDI pour (Belief, Desire, Itention) représentant les agents les plus intelligents par rapport aux trois autres types d'agents.

## 3. Les Systèmes Multi Agents (SMA)

Depuis les années 80, les systèmes multi-agents ne cessent de se développer et d'étendre leur domaine d'application. Ils constituent une des disciplines les plus actives actuellement dans plusieurs domaines en particulier l'intelligence artificielle, les systèmes distribués et le génie logiciel [BEN, 15].

### 3.1. Définition

Un système multi-agent (SMA) est : *« un ensemble d'agents évoluant au sein d'un même environnement. Il fournit des moyens de communication aux agents, qui interagissent afin d'effectuer les tâches globales du système »* [TAG, 11].

L'approche Systèmes Multi-Agents permet d'appréhender, de modéliser, de simuler des systèmes complexes, c'est-à-dire des systèmes constitués d'un nombre d'entités important en interaction entre eux et avec le monde extérieur. Les agents sont autonomes et interagissent via un environnement. Ils doivent être en mesure de fournir des solutions robustes et être capables de s'adapter dans des environnements qui peuvent être imprévisibles [YAN, 12], la figure suivante (Figure II.1) illustre les agents et leurs interactions avec leurs environnement.

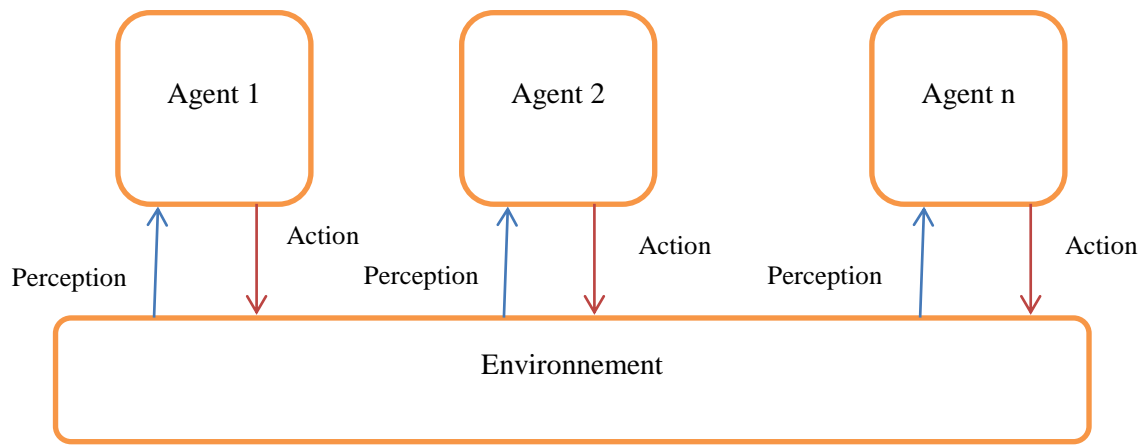

**Figure II.1 :** Les agents et leur environnement.

**Jack FERBER** [FER, 95] rappelle qu'un système multi-agents (ou SMA) est un système composé des éléments suivants:

- **Un environnement E**, c'est-à-dire un espace disposant généralement d'une métrique.
- **Un ensemble d'objets O**. Ces objets sont situés, c'est-à-dire que, pour tout objet, il est possible, à un moment donné, d'associer une position dans E. Ces objets sont passifs, c'est-à-dire qu'ils peuvent être perçus, créés, détruits et modifiés par les agents.
- **Un ensemble A d'agents**, qui sont des objets particuliers (A $\subseteq$ O), lesquels représentent les entités actives du système.
- **Un ensemble de relations R** qui unissent des objets (et donc des agents) entre eux.

- **Un ensemble d'opérations Op** permettant aux agents de A de percevoir, produire, consommer, transformer et manipuler des objets de O.
- **Des opérateurs** chargés de représenter l'application de ces opérations et la réaction du monde à cette tentative de modification, que l'on appellera les lois de l'univers.

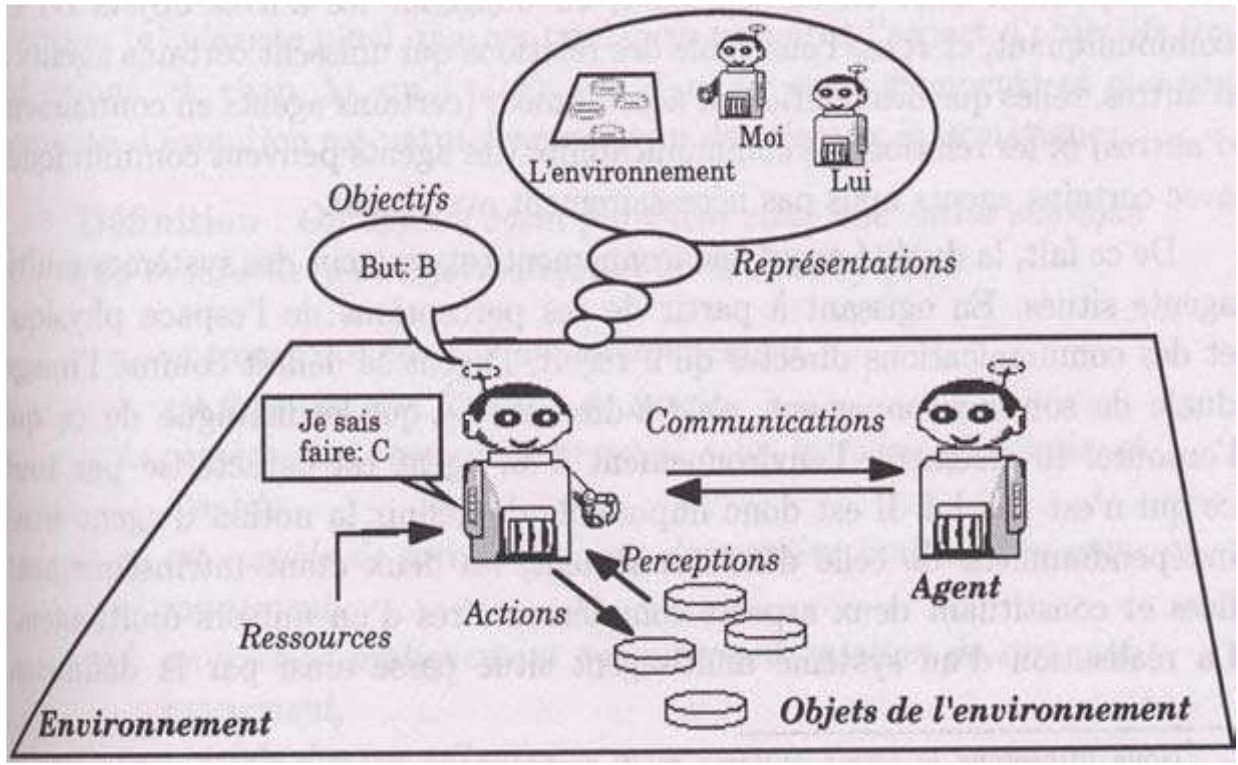

**Figure II.2 :** Représentation d'un agent en interaction avec son environnement et les autres agents.

## 4. L'interaction entre les agents

Placer un ensemble des agents dans un même environnement ne définit pas nécessairement un SMA. Les agents doivent pouvoir interagir entre eux, se coordonner et éventuellement coopérer [ABD, 12].

## 5. Définition de l'interaction

Selon Jack FERBER [FER, 95] l'interaction est définie comme suit :

« *Une interaction est la mise en relation dynamique de deux ou plusieurs agents par le biais d'un ensemble d'actions réciproques. Les interactions sont non seulement la conséquence d'actions effectuées par plusieurs agents en même temps, mais aussi l'élément nécessaire à la constitution d'organisations sociales* »

Autrement dit, l'interaction ne peut pas être faite que s'il y a une influence d'un agent par un autre ou par son environnement.

### 5.1. Les modes d'interactions entre les agents

Dans les systèmes multi-agents, la communication est souvent l'un des moyens utilisés pour échanger des informations entre agents (ex. plans, résultats partiels, buts, etc.).

Généralement, deux modes sont cités [ABD, 12] :

- **Direct :** correspond à des échanges entre les agents sans passer par l'environnement.
- **Indirect :** correspond à une interaction entre les agents et l'environnement.

### 5.2. Les types des interactions entre les agents

Parmi les types d'interactions entre les agents, on cite [ABD, 12] :

- La **coopération :** les agents travaillent ensemble pour la résolution d'un but commun ;
- La **coordination :** organiser la résolution d'un problème de telle sorte que les interactions nuisibles soient évitées ou que les interactions bénéfiques soient exploitées ;
- La **négociation** : parvenir à un accord acceptable pour toutes les parties concernées.

## 6. Les langages de communication entre les agents

Les ACLs (Agent Communication Language) sont basés sur la linguistique et principalement la théorie des Actes de Langage. Il s'agit de permettre l'échange d'informations entre les entités intelligentes et de faciliter l'interopérabilité entre les différentes plateformes multi-agents. Ces langages font donc abstraction du mode de transport et sont codés sous forme de chaines de caractères ASCII [PIE, 05].

Un échange de connaissances ne se contente que rarement d'un seul message, en général il est nécessaire de s'en échanger plusieurs : une question suivie d'une réponse par exemple.

Une conversation entre deux agents correspond à une suite de messages qu'ils s'échangent pour atteindre le but qui a initié le dialogue. Dans le cas d'agents ayant des avis et intentions sur le monde qui les entourent, on pourrait laisser ces intentions guider le choix du message à envoyer [PIE, 05]. Il existe principalement deux langages spécifiques : KQML et FIPA ACL .

### 6.1. KQML (Knowledge Query and Manipulation Language)

KQML est un ACL développé au début des années 90, par l'équipe Externat Interfaces Working Group du projet DARPA9-KSE10 [PIE, 05].

KQML est un langage issu des travaux de la knowledge sharing effort. Il s'agit d'un langage qui vise à définir un ensemble d'actes de langages qui soit standard et utile. Ces actes de langage appelés aussi performatifs, sont utilisés par les agents pour échanger des informations. La forme de base du protocole est [BEN, 15] :

KQML-performative

Sender <word>

Receiver <word>

Language<word>

Ontology<word>

Content <word>

Le but de KQML était principalement de fournir un langage de communication basé sur les actes de langage et indépendant des applications qui l'utilisent [PIE, 05].

### 6.2. FIPA ACL

FIPA ACL est, comme son nom l'indique, un projet de la FIPA pour fournir un ACL simplifiant l'interopérabilité entre les différentes plateformes multi-agents [PIE, 05].

L'ACL de la FIPA ressemble beaucoup à KQML, mais il comporte moins de performatifs que KQML. De plus, il est défini de façon beaucoup moins ambiguë.

## 7. Les tableaux noirs

En intelligence artificiel, la technique de tableaux noir est très utilisée pour spécifier une mémoire partagée par divers systèmes. Le principe du tableau noir appelé BLACK BOARD est basé sur deux éléments [BEN, 15] :

1- Un ensemble d'entités qui sont sources de savoir, chacune étant spécialisée dans un type de connaissances.
2- Un espace structuré de partage des données utilisé par les entités pour communiquer.

Un blackboard est simplement une structure de données partagées entre divers agents. Ces derniers peuvent la consulter pour obtenir des informations sur l'état actuel du problème ou y écrire la partie de la solution qu'ils ont obtenue.

Dans un SMA utilisant un tableau noir, les agents peuvent écrire des messages, insérer des résultats partiels de leurs calculs et obtenir de l'information. Le tableau noir est, en général, partitionné en plusieurs niveaux qui sont spécifiques à l'application. Les agents qui travaillent sur un niveau particulier peuvent accéder aux informations contenues dans le niveau correspondant du tableau noir ainsi que dans des niveaux adjacents [CHA et al, 01].

La Figure suivante représente une architecture d'un SMA à base de BlackBoard :

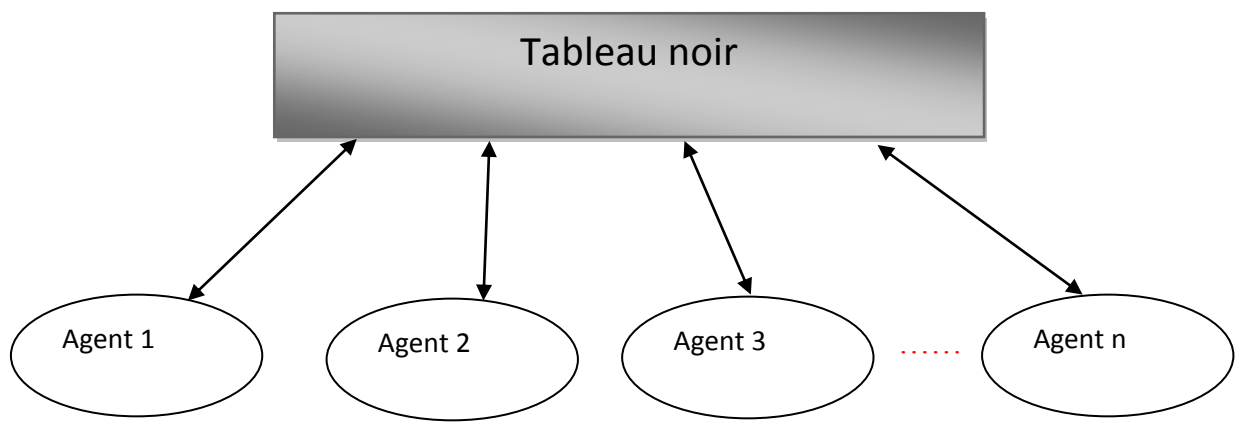

**Figure II. 3:** Architecture d'un Système multi agent à base de blackboard

## 8. La négociation dans les SMA

La négociation est utilisée dans le domaine du commerce électronique notamment en utilisant les enchères, dans le domaine des télécommunications par exemple pour partager une bande passante et dans les systèmes multi-agents (SMA) pour l'allocation de tâches et de ressources [HEL, 04], et elle fait partie de beaucoup de travaux de recherches dans différents domaines qui utilisent les agents ou des systèmes multi-agents tel que l'IA.

### 8.1. Définition

Par négociation, nous entendons une discussion dans laquelle les individus échangent des informations et arrivent à un accord, c'est le processus à travers duquel plusieurs entités prennent une décision commune. Ces individus expriment d'abord des demandes contradictoires pour essayer de trouver par la suite un accord satisfaisant par concession ou par la recherche de nouvelles alternatives [HAM, 08].

## 9. Les formes de la négociation

La négociation dans les SMA prend plusieurs formes selon le type du système multi-agents et selon le processus de décision utilisé. Parmi ces formes, on a :

### 9.1. La négociation par le Vote

Les systèmes de votes sont utilisés pour élire une alternative parmi déférentes alternatives possibles. Cela revient à proposer l'alternative et à recueillir les votes pour et les votes contre, cette alternative. Les systèmes complexes impliquent un nombre d'alternatives supérieur à deux, les votants devant alors choisir l'alternative qu'ils préfèrent [BOU, 12].

Pour cette forme, on trouve plusieurs méthodes, mais on liste:

- **La méthode de Borda:** proposée par Jean Charles Borda en 1781. Dans cette méthode, l'idée générale est d'attribuer des points à chaque alternative, le principe est le suivant [OUF, 09] :

   Pour chaque liste de préférences, l'alternative classée première remporte (n-1) points, la seconde (n-2) et ainsi de suite jusqu'à la dernière qui obtient 0 point. Pour chaque alternative, on calcule le nombre total des points qu'elle a recueilli, l'alternative ayant le meilleur score est déclarée le choix social [OUF, 09].

### 9.2. La négociation par les enchères

Cette forme de la négociation est la plus utilisée en internet, pour les sites de vente par exemple les véhicules, les objets d'arts, les produits périssables, et un certain type de jeux comme le bridge.

Elles permettent à des personnes de vendre leurs biens au meilleur prix possible. Chacune de ces enchères implique un vendeur et une assemblée d'acheteurs, mais le déroulement de l'enchère diffère selon le type utilisé. Le vendeur fixe son prix de départ et son prix de réserve, c'est-à-dire le plus petit prix acceptable pour vendre le bien [BEN, 15]. Les principales formes d'enchères sont les suivantes [OUF, 09]:

1. **Enchère anglaise (premier-prix offre-publique) :** Chaque participant annonce publiquement son offre. Le participant avec la plus grande soumission gagne l'objet au prix de son offre.

2. **Enchère hollandaise (descendante) :** L'initiateur diminue tout le temps le prix jusqu'à ce qu'un des participants achète l'objet au prix actuel.

3. **Enchère premier-prix offre-cachée :** Chaque participant soumet une offre sans savoir les offres des autres. Celui qui fait la plus grande soumission gagne l'objet et paye le montant de son offre.

4. **Enchère de Vickery (deuxième-prix offre-cachée) :** Chaque participant soumet une offre sans savoir les offres des autres. Celui avec la plus grande offre gagne, mais au prix de la deuxième plus grande offre.

### 9.3. La négociation par l'argumentation

La négociation à base d'argumentation est utilisée chez des agents possédant une base de connaissances avec des prédicats et des règles d'inférences. L'argumentation a pour but de modifier les croyances des autres agents afin qu'ils adoptent le même point de vue, les mêmes croyances ainsi les même intensions que l'agent argumentant [HAM, 08].

La négociation par argumentation c'est le fait d'utiliser des Arguments pour convaincre l'autre d'accepter la proposition faite, il existe différents types d'arguments, chaque type d'argument définit des pré-conditions pour son utilisation, si elles sont remplies, alors l'agent peut utiliser l'argument. L'agent a besoin d'une stratégie pour décider quel argument utiliser.

- **Appels à une promesse passée :** le négociateur A rappelle B d'une promesse passée concernant le NO, i.e., l'agent B a promis dans une négociation passée à l'agent A d'offrir ou effectuer un NO. Pré-conditions A doit vérifier si une promesse d'un NO a été reçue dans passé dans une négociation conclue avec succès.
- **Promesse d'une récompense future :** le négociateur A promet de faire NO pour un autre agent B à un moment dans le futur. Pré-conditions A doit trouver un désir de l'agent B pour un moment dans le futur, si possible un désir qui peut être satisfait par une action (service) que A peut effectuer mais B non.
- **Appels au propre intérêt :** l'agent A croit que arrivant à un accord sur NO est dans l'intérêt de B et essaye de convaincre B de ça. Pré-conditions A doit trouver (ou inférer) un des désirs de B qui sera satisfaite si Ba NO ou A doit trouver un autre objet de négociation NO qui a été offert auparavant dans le marché et il croit que NO est mieux que NO.
- **Menace** le négociateur menace de refuser faire/offrir quelque chose à B ou il menace qu'il fera quelque chose qui contredit les désirs de B. Pré-conditions A doit trouver un des désirs de B directement satisfaite par un NO que A peut offrir ou A doit trouver une action qui est contradictoire avec ce qu'il croit être un des désir de B

La négociation basée sur l'argumentation est très proche de la négociation humaine, mais elle n'a pas encore atteint un stade de maturité. Même en exploitant les capacités d'inférence des langages de programmation basés sur la logique, le problème d'implémenter une négociation à base d'augmentation reste très difficile. Peu sont les programmes qui ont fait avec sucées ce genre de négociation [HAM, 08].

### 9.4. La négociation par Protocole de Contrat-Net (CNP)

Le protocole Contra-Net est un protocole proposé par Smith en 1980, c'est un protocole de haut niveau pour la communication entre les nœuds d'un résolveur de problèmes distribué.

En effet, l'initiateur lance un appel d'offres pour la réalisation d'une tâche, le message est envoyé à tous les agents qu'il estime capables de réaliser cette tâche. A partir de la description de la tâche, les agents participants construisent une proposition qu'ils envoient à l'agent initiateur (manager) ; A cet effet, le manager reçoit et évalue les propositions et attribue la tâche au meilleur contractant, le contractant auquel la tâche est confiée envoie un message au manager en lui confirmant son intention de l'accomplir [HAM, 08].

Le modèle Contra-Net étendu est apparue comme une extension du CNP, il est basé sur une décomposition décentralisée de tâches, l'agent gestionnaire(initiateur) fractionne une tâche en plusieurs sous tâches et les annonces à un agent ou un groupe d'agents contractants mais contrairement au CNP où une tâche ne peut être qu'accordée ou rejetée, dans le Contra-Net étendu, une tâche peut être accordée temporairement, rejetée temporairement, accordée définitivement, rejetée définitivement par les agents [OUF, 09].

### 9.5. La négociation par Concession Monotone

Ce protocole est basé sur l'allocation des tâches (on trouve aussi des ressources), et le problème d'allocation des tâches a été étudié par Rosenschein et Zlotkinen en 1994.

❖ **Principe**

Le processus commence lorsque chaque agent propose une offre qui maximise son utilité. Puis itérativement, les agents font des concessions jusqu'à arriver à une entente.
Ce protocole est décrit comme suit : Les deux agents partent du point de rencontre. A chaque étape, les deux agents font une proposition en termes d'utilité. Si cela les satisfait alors la négociation s'arrête, sinon l'un deux (ou les deux) doit concéder en offrant à l'autre une meilleure utilité (il diminue par conséquent la sienne). Ce protocole se termine après un nombre fini des tours.

Ce protocole sera détaillé dans le chapitre suivant et fait l'objet de notre étude.

### 9.6. Les autres formes de négociation

Il existe d'autres formes de négociation moins connues du grand public, ces négociations sont tout aussi nombreuses et variées. Parmi ces négociations, on retrouve, principalement, le take it or leave it offer et les négociations multi attributs [OUF,09].

❖ **Le take it or leave it offer**

Cette forme de négociation est très primaire, puisqu'elle consiste à formuler une proposition qui est à prendre ou à laisser par le ou les participants. Cette négociation se déroule sur un seul tour, sans contre-proposition ni renégociation. C'est celle que l'on rencontre tous les jours pour acheter son pain, par exemple. Elle ressemble aux offres scellées au meilleur et au second meilleur prix, la différence vient du fait que dans les enchères, le prix est proposé par les acheteurs, tandis que dans le *take it orleave it* offer, c'est le vendeur qui propose un prix ferme. Le protocole est donc très simple : le vendeur propose sa ressource (bien, service, etc.) à un prix ferme à qui soit accepté, soit refuse. S'il accepte, l'acheteur paie le prix pour obtenir la ressource. Le take it or leave it offer ne présente pas pour beaucoup de chercheurs une véritable négociation, et nous sommes d'accord avec eux [OUF, 09].

❖ **Les négociations multi-attributs**

Les négociations multi-attributs, comme leur nom l'indique, sont des négociations qui impliquent différents attributs devant être négociés. Elles sont directement opposées aux enchères qui n'impliquent qu'un seul attribut : un prix. Cette forme de négociation est cependant très répandue et à la base de nombreuses variantes de négociation. Un exemple de négociation multi-attributs est la négociation d'une voiture chez un concessionnaire. Le modèle, la motorisation, la couleur et de nombreuses options comme la climatisation, la direction assistée ou encore la présence d'airbags, en plus du prix, seront négociés**.** [OUF, 09].

## 10. Conclusion

La négociation est un processus qui permet de résoudre un conflit considérable existant entre les agents dans un SMA pour arriver à une solution satisfaisante, et acceptable par tous les agents, elle peut prendre plusieurs formes.

Ce chapitre constitue une présentation de la négociation et des méthodes de négociation entre les agents dans les systèmes multi agents.

Dans le chapitre suivant, nous détaillons les aspects les plus importants de la conception et la modélisation de notre étude.

# Chapitre III

# Conception et Modélisation

## 1. Introduction

Dans les chapitres précédents, nous avons abordé l'aspect théorique des différents concepts liés à notre problématique: l'aide à la décision multicritères et multi décideurs.

Dans ce chapitre, nous allons décrire d'une façon détaillée le système d'aide à la décision de groupe proposé. Ce dernier intègre plusieurs variantes qui contribuent aux mieux pour analyser le contexte de notre étude. Il repose sur une utilisation combinée de méthodes d'analyse multicritères et des techniques de négociation.

Tout au long de ce chapitre, nous allons présenter les détails relatifs à la conception et la modélisation de notre contribution.

## 2. Objectifs visés

Nous nous intéressons, dans cette étude, à l'élaboration d'un système d'aide multicritères à la décision de groupe. Le système proposé assure la négociation ainsi que la participation d'une diversité de décideurs géographiquement dispersés en vue d'aboutir à une décision de groupe et un accord mutuellement acceptable.

L'approche décisionnelle adoptée prend appui, principalement, sur l'utilisation d'un couplage SMA-SIAMD (SMA : Système Multi Agents, SIAMD : Système Interactif d'Aide MultiCritères à la Décision).

Le module SMA est responsable de la représentation des différents décideurs impliqués, tandis que, le module SIAMD est chargé d'exprimer les préférences de chaque décideur et générer, en conséquence, la décision de groupe après un processus de négociation.

Les objectifs spécifiques de cette étude sont comme suit :
1- Aider les décideurs à exprimer leurs préférences.
2- Modéliser la problématique décisionnelle par une approche basée agents.
3- Proposer un protocole de négociation basé sur la concession monotone.
4- Intégrer des méthodes d'analyse multicritères par agrégation totale AHP et par agrégation partielle Prométhée dans le processus de négociation.
5- Proposer une interface Web pour le GDSS.

## 3. Formulation de notre problème décisionnel

Le problème décisionnel abordé est multicritères et multi-décideurs exprimé par le modèle suivant (A, F, MP, PS, ND, Poids_décideur)

- **A :** désigne l'ensemble des actions proposées (solutions possibles au problème).
- **F :** Famille des critères identifiée.
- **MP :** Matrice de performance, ou matrice d'évaluations des actions par rapport aux critères, c'est-à-dire l'ensemble des vecteurs de performances.
- **PS :** Paramètres subjectifs exprimés par le décideur.
- **ND** : le nombre des décideurs, ou le nombre des participants à la résolution du problème.
- **Poids_décideur :** le poids de chaque décideur dans la décision de groupe.

## 4. Architecture globale de notre système

Le système d'aide à la décision de groupe (GDSS) proposé et mis en œuvre est conçu en utilisant un système multi agents (SMA) doté d'un protocole de négociation exploitant deux méthodes multicritères d'aide à la décision.

### ❖ Le module Système multi agents (SMA)

Un système multi agents (SMA) est définit dans la littérature comme étant « un système où plusieurs agents au moins partiellement autonomes interagissent entre eux pour réaliser des objectifs communs ». Cette définition met l'accent sur l'importance des SMA et le rôle qu'ils jouent en grande partie dans la modélisation de problèmes informatiques divers, malgré cette diversité, ces problèmes ont en commun la présence de plusieurs agents qui collaborent entre eux.

Dans le contexte de notre étude et afin de traiter un problème décisionnel de groupe qualifié de complexe, les SMA sont utilisés pour modéliser l'interaction entre les différents décideurs. A ce titre, deux types d'agents sont à considérer : **les agents participants** qui sont les décideurs et **un agent initiateur** (coordinateur, facilitateur, médiateur) qui gère la communication et la négociation entre les différents décideurs.

- Notre architecture est destinée pour un groupe de décideurs, où chaque décideur doit interagir avec le système à travers une interface web. Cette dernière joue le rôle du coordinateur du groupe.
- Le décideur participant doit s'inscrire pour pouvoir participer à la négociation, il s'identifie en introduisant ses coordonnées (Nom, Prénom) ainsi que son poids, puis il exprime ses préférences ou paramètres subjectifs (Indifférences, préférences, poids entres les critères, échelle de Saaty pour les critères, échelle de Saaty pour les Actions).
- l'initiateur lance, par la suite, la négociation, chaque participant fait appel à une méthode multicritères pour effectuer le rangement soit par agrégation totale par appel à la méthode AHP si le nombre des critères identifiés est inférieur à 10 ; soit partielle par appel à la méthode Prométhée si le nombre des critères est supérieur à 10.
- L'initiateur contrôle la négociation jusqu'à trouver la solution finale qui sera communiquée aux participants.

La figure suivante illustre une vue d'ensemble du système proposé.

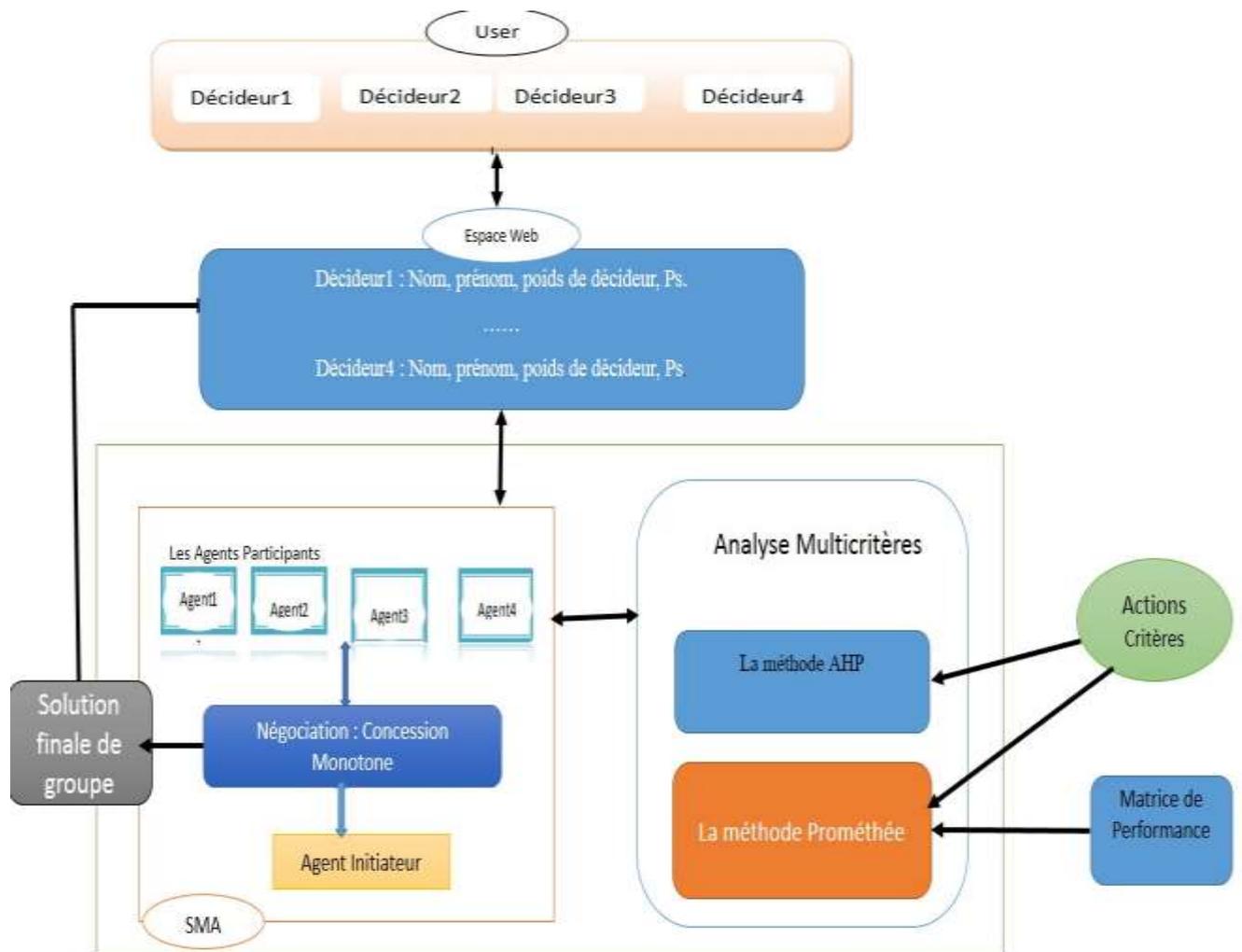

**Figure III.1** : Architecture générale de notre système.

## 5. L'architecture fonctionnelle de notre système

La démarche décisionnelle adoptée par le GDSS est donnée par l'organigramme suivant :

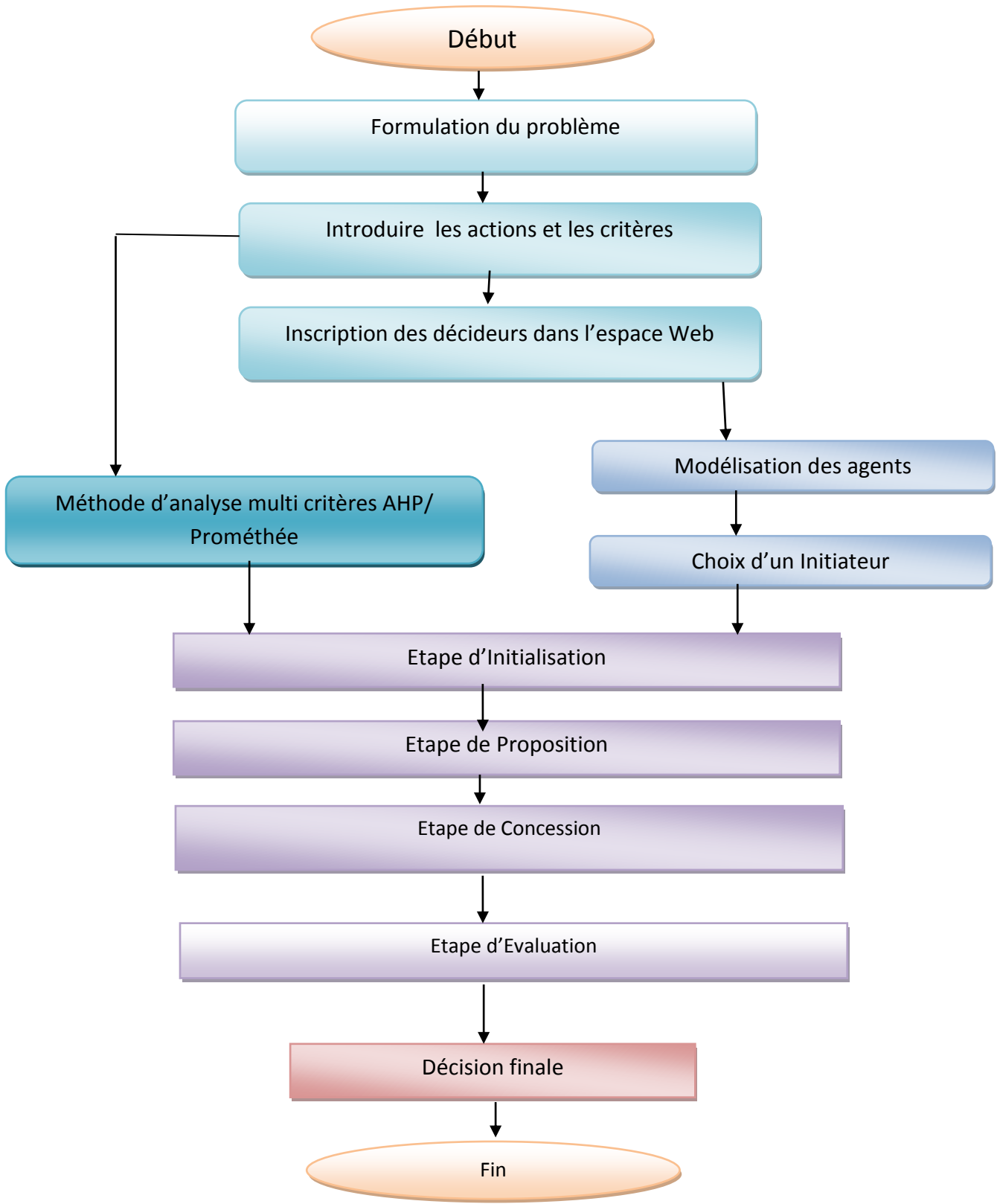

**Figure III.2 :** Démarche décisionnelle adoptée par le GDSS.

## 6. Les étapes du protocole de négociation

Notre protocole passe par cinq étapes principales décrites dans ce qui suit :

1. **Etape d'initialisation :** Cette étape marque le début de la négociation, les participants sont appelés à exprimer leur préférences (à travers le message **Request**) concernant les actions, chaque participant effectue un classement des ressources de la meilleure à la moins bonne par un appel à une méthode multicritères soit la méthode AHP, soit la méthode Prométhée.
2. **Etape de proposition :** Après que l'initiateur reçoit tous les classements des actions de la part des participants, il propose un contrat (proposition) concernant une ressource donnée (à travers le message **Propose**), et il l'émet aux participants, ces derniers vont soit accepter ce contrat, soit concéder soit le refuser selon leur vecteur de préférences (classement des actions).
3. **Etape de concession:** Après la réception d'une proposition, les participants doivent diviser leur vecteur de rangement en 3 intervalles,
   - si la proposition reçue se trouve dans la première partie de l'intervalle alors il accepte la proposition et envoie le message **Accept** à l'initiateur,
   - si elle se trouve dans la 2ème partie, alors il fait une concession et il envoie un message **Conceed** à l'initiateur,
   - sinon il refuse et envoie un message **Refuse** à l'initiateur.
4. **Etape d'évaluation :** Lorsque l'initiateur reçoit toutes les réponses des participants concernant une proposition de contrat, il comptabilise le nombre d'agents participants ayant accepté sa proposition,
   - si ce nombre est supérieur ou égal à un certain seuil (seuil de négociation) alors la négociation est un succès ;
   - sinon l'initiateur doit faire une modification du contrat proposé en s'inspirant des rangements des agents, puis il revient à l'étape de proposition.
5. **Etape de décision :** C'est la dernière étape du protocole suggéré. C'est la fin de la négociation, une décision est prise par l'initiateur selon les réponses des participants aux propositions qu'il leur a fait, et l'initiateur confirme la décision trouvé, et il l'envoie à tous les participants à travers le message **Confirm.**

## 7. Le langage du protocole de négociation

Le langage de la négociation entre les agents dans le protocole proposé s'articule autour d'un ensemble de primitives de l'Initiateur et d'autres des participants.

Le tableau suivant récapitule les primitives spécifiques à l'initiateur.

| Primitives d'Initiateur | Explication |
|---|---|
| **Request()** | Demande aux participants d'envoyer le rangement total des actions. |
| **Propose()** | Envoie un contrat aux participants, généré au niveau de l'étape de proposition |
| **Confirm()** | Envoie la solution finale aux participants. |

**Tableau III.1 :** Les primitives de l'Initiateur.

Le tableau suivant récapitule les primitives spécifiques aux participants.

| Primitives d'un Participant | Explication |
|---|---|
| **Inform()** | Envoie à l'initiateur le classement des ressources selon les préférences des agents. |
| **Accept()** | Accepte la proposition |
| **Conceed()** | Fait une concession |
| **Refus()** | Refuse la proposition |

**Tableau III.2 :** Les primitives d'un Participant.

## 8. Le Seuil de négociation

Le seuil de négociation est le pourcentage des participants à partir duquel la négociation est jugée comme réussie et s'arrête par ce fait.

## 9. Les ressources de négociation

Dans le contexte de notre étude, les ressources de négociation sont les actions partagées entre les agents participants, et l'initiateur.

## 10. La méthode AHP

La méthode AHP est une méthode développée par **Thomass Saaty** dans les années de 1980 [HAM, 16], cette méthode est une méthode par agrégation totale (Annexe A). C'est un outil efficace pour faire face à la prise de décision complexe, et peut aider le décideur à prendre la meilleure décision.

La méthode AHP consiste à représenter un problème de décision par une structure hiérarchique reflétant les interactions entre les divers éléments du problème, à procéder ensuite à des comparaisons par paires des éléments de la hiérarchie, et enfin à déterminer les priorités des actions [HAM, 03].

La résolution de problèmes décisionnels selon la méthode AHP repose sur trois principes fondamentaux :

a) **Décomposition :** Le principe de la décomposition est appliqué pour structurer un problème complexe en une hiérarchie de critères, sub-critères, subsub critères ainsi de suite.

b) **Comparaison des jugements :** Le principe de la comparaison des jugements est utilisé pour construire des comparaisons par paire de toutes les combinaisons d'éléments du même niveau on respectant le père de chaque niveau .Ces comparaisons sont utilisés pour dériver les priorités locales des éléments dans leur niveau.

c) **La synthèse des propriétés ou composition hiérarchique :** Le principe de la synthèse des propriétés est de multiplier les priorités locales de chaque niveau par les priorités globales du niveau père pour obtenir les priorités globales de toute la hiérarchie par la suite on ajoute les priorités globales du niveau le plus bas (généralement Ce sont les alternatives).

### a. Etapes de la méthode AHP

La méthode AHP passe par plusieurs étapes, elle procède comme suit:

- **Etape 01 :** Décomposer le problème en une hiérarchie d'éléments inter-reliés. Au sommet de la hiérarchie, on trouve l'objectif, et dans les niveaux inférieurs, les éléments contribuant à atteindre cet objectif. Le dernier niveau est celui des actions [HAM, 03].

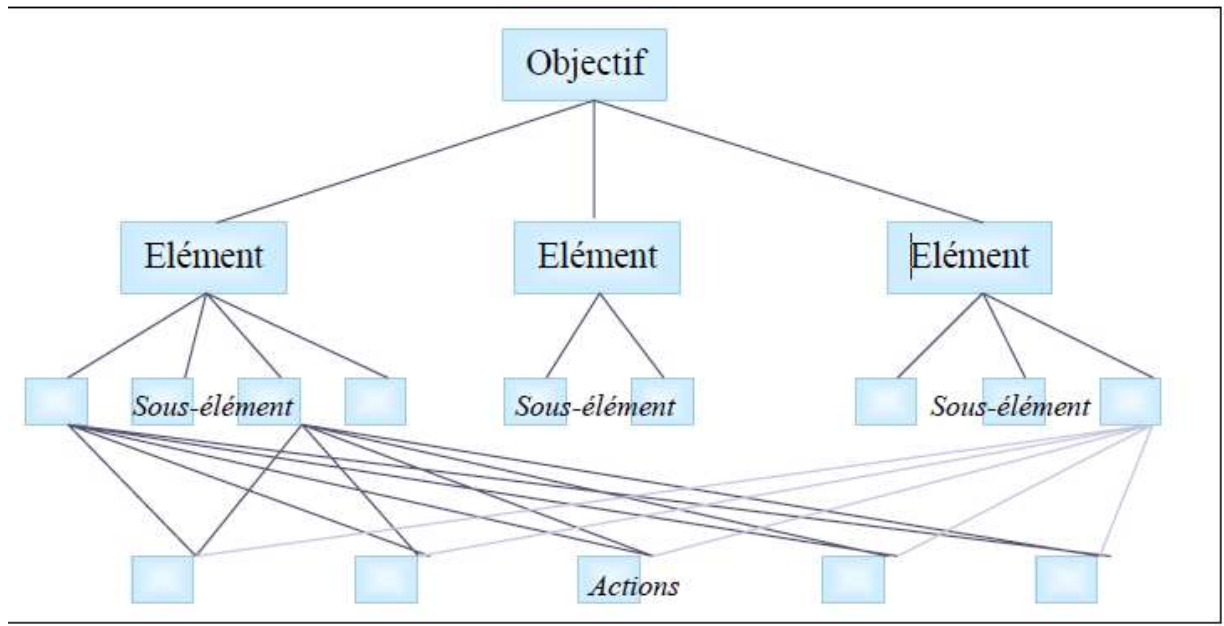

**Figure III.3 :** Structure hiérarchique d'un problème selon la méthode AHP.

- **Etape 02 :** Procéder à des comparaisons par paires des éléments de chaque niveau hiérarchique par rapport à un élément du niveau hiérarchique supérieur. Cette étape permet de construire des matrices de comparaisons. Les valeurs de ces matrices sont obtenues par la transformation des jugements en valeurs numériques selon l'échelle de Saaty (Echelle de comparaisons binaires), tout en respectant le principe de réciprocité [HAM, 03] :

$Pc(EA,EB) = \frac{1}{Pc(EB,EA)}$

- **Etape 03 :** Déterminer le pourcentage des critères, par l'utilisation le matrice de comparaison entres les critères.
- **Etape 04 :** Déterminer le pourcentage des actions pour chaque critère, par l'utilisation des matrice de comparaison entres les actions pour un critère.
- **Etape 05 :** Trier le vecteur trouvé par la multiplication de la matrice d'Etape 04 et vecteur d'Etape 03. puis, on fait le choix de la meilleure solution.

La figue suivante illustre plus précisément les étapes à suivre dans l'application de cette méthode :

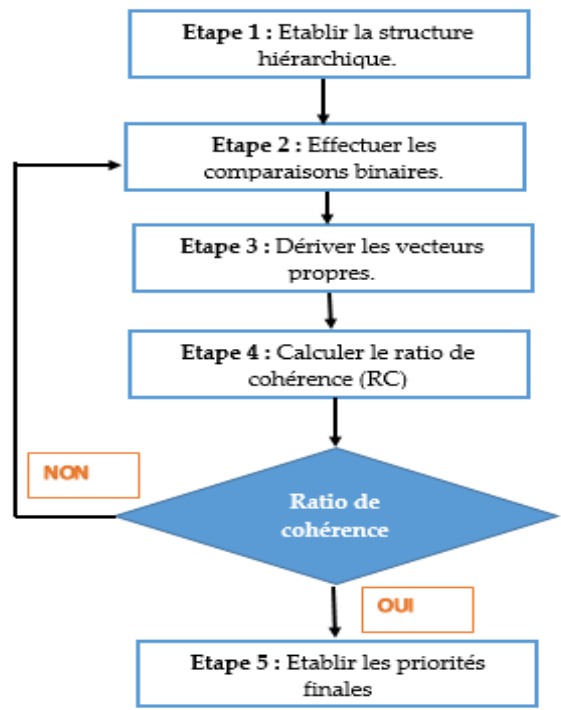

**Figure III.4 :** La démarche d'utilisation de la méthode AHP

### b. Les avantages de la méthode AHP

Parmi les avantages, on cite [ALN, 16] :

- Modélisation du problème par une structure hiérarchique.
- Utilisation d'une échelle sémantique exprimant les préférences du décideur.
- Grande flexibilité.
- Ne necessite pas une matrice de performances.

### c. Les limites de la méthode AHP

Parmi les limites, on retrouve [ALN, 16]:

- Trop lourde en nombre de comparaisons par pairs à effectuer pour un grand nombre de critères.
- L'ordre de priorité des actions est influencé par l'ajout ou par la suppression d'une ou plusieurs actions (Renversement de rang).
- Comparaisons délicates des critères et des actions.

## 11. La méthode Prométhée II

PROMETHEE II est une méthode multicritères qui permet de résoudre la problématique de rangement afin de classer l'ensemble des actions potentielles de la meilleure vers la pire (préordre total).

C'est une méthode procédant par agrégation partielle et utilise des relations de surclassement [HAM, 16]..

### a. Principe de la méthode Prométhée II

Elle procède comme suit :

- ❖ Choix d'un critère concernant l'intensité de préférence

Dans notre étude, nous avons choisi le critère mixte.

- ❖ Calcul de l'intensité de préférence

$p(d) = 0$ si $d \leq q_j$, $p(d) = (d - q_j) / (p_j - q_j)$ si $q_j < d \leq p_j$ et $p(d) = 1$ sinon

« a » et « b » étant deux actions potentielles,

« d » est la différence entre la performance de « a » et la performance de « b » ($g_j(a) - g_j(b)$).

$q_j$ est le seuil d'indifférence, et $p_j$ est le seuil de préférence.

- ❖ Calcul de l'Indicateur de préférence

$\pi(a, b) = \Sigma W_j * P_j(a, b) / \Sigma W_j$

$W_j$ étant le poids du critère j.

- ❖ Calcul des Flux sortant et entrant

$\phi^+(a) = \Sigma \pi(a, x)$     et     $\phi^-(a) = \Sigma \pi(x, a)$

- ❖ Calcul des flux globaux

$\phi(a) = \phi^+(a) - \phi^-(a)$.

L'organigramme suivant illustre, en détails, la démarche d'utilisation de PROMETHEE II.

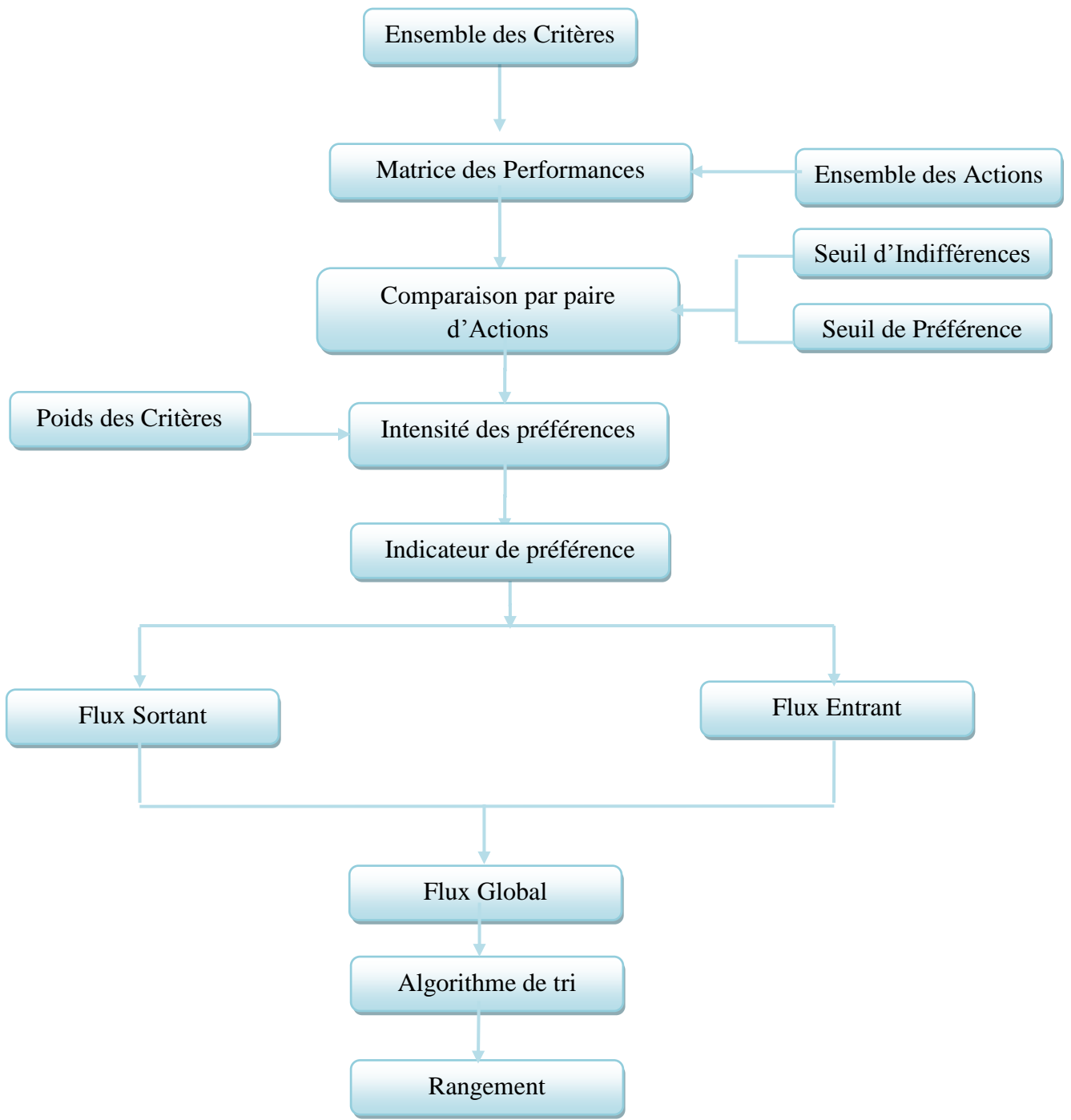

**Figure III.5** : Démarche d'utilisation de PROMETHEE II.

*Comment le coordinateur génère une proposition à un tour t ?*

A chaque tour, le contrat (proposition) est généré en s'inspirant des rangements effectués par les participants. En effet, le coordinateur associe un score à chaque ressource en prenant en considération le poids de l'agent participant ainsi que le rang de la ressource. Pour calculer le score de chaque ressource $R_i$ (i= 1,.. n) lors d'un certain tour t, nous avons utilisé la formule suivante :

SCORE($R_i$)=∑ POID(participant[j])*RANG(Ri, participant[j])   ( j=1,n  n, m : nombre de ressources et   participants respectivement)

Avec :

POID (participant[j]) : compte tenu que dans la réalité, les représentants politiques, par exemple, n'ont pas le même poids que les associations de protection de l'environnement lors d'une décision en AT, nous associons à chaque participant j un poids différent.

RANG ($R_i$, participant[j]) : le rang associé à la ressource (action) i par le participant j dans son vecteur de préférence (rangement fourni par AHP ou Prométhée) ;

Comme dans les méthodes de scorages, la ressource qui a obtenu le score le plus bas sera la ressource gagnante et le coordinateur la proposera dans le nouveau contrat.

## 12. Structure des fichiers des données utilisées

Nous avons utilisé des fichiers textes pour enregistrer nos données, et les transmettre entre page Web et le SMA, nous disposons de 3 types de fichiers.

1- Un fichier pour sauvegarder les paramètres subjectifs de chaque décideur utilisé par Prométhée.
2- Un fichier pour sauvegarder les paramètres subjectifs d'un décideur utilisé par AHP.
3- Un fichier texte pour sauvegarder tous les décideurs inscrits et membres du groupe de décision, ainsi que leurs poids.

## 13.Modélisation UML

UML est l'abréviation de Unfied Modeling Language, c'est-à-dire langage unifié pour la modélisation. C'est une notation graphique destinée à la création de modèles orienté objet en vue de l'analyse et de la modélisation de logiciel orienté objet. Ce n'est pas une méthode, c'est un ensemble d'outils permettant la modélisation de la future application informatique [UML 2, 09].

Il existe plusieurs diagrammes d'UML, nous avons choisi de modéliser notre système en utilisant quatre diagrammes :

1. Diagramme de classe.
2. Diagramme de cas d'utilisation.
3. Diagramme de séquences.
4. Diagramme de Collaboration.

Le logiciel de la modélisation utilisé est le **StarUML** de version 2.8.1, c'est un logiciel de modélisation UML, cédé comme open source par son éditeur, à la fin de son exploitation commerciale, sous une licence modifiée de GNU GPL, StarUML il gère la plupart des diagrammes spécifiés dans la norme UML 2.0 [Net, 1].

- **Le diagramme de classe**

Avant d'établir le diagramme de classe, il faut de faire définir un dictionnaire de données qui permet de classer et trier les informations, et les coder. Notre dictionnaire de donnée est le suivant (voir Tableau III.3) :

| Code | Définition |
|---|---|
| Id_deci | Identifiant de la classe décideur |
| Consulter_Trace_Neg | Consulter la trace de la négociation |
| Id_pag | Identifiant de la classe pageWeb |
| Nbr_decideur | Nombre des décideurs |
| Id_Agent | Identifiant de la classe Agent |
| Id_AHP | Identifiant de la classe AHP |
| Id_Prom | Identifiant de la classe Prométhée |
| Id_Init | Identifiant de la classe Initiateur |
| Id_Part | Identifiant de la classe Participant |
| Nom_P | Nom d'agent participant |

**Tableau III.3** : Dictionnaire des données pour le diagramme de classe

La figure III.3 illustre le diagramme de classe de notre système.

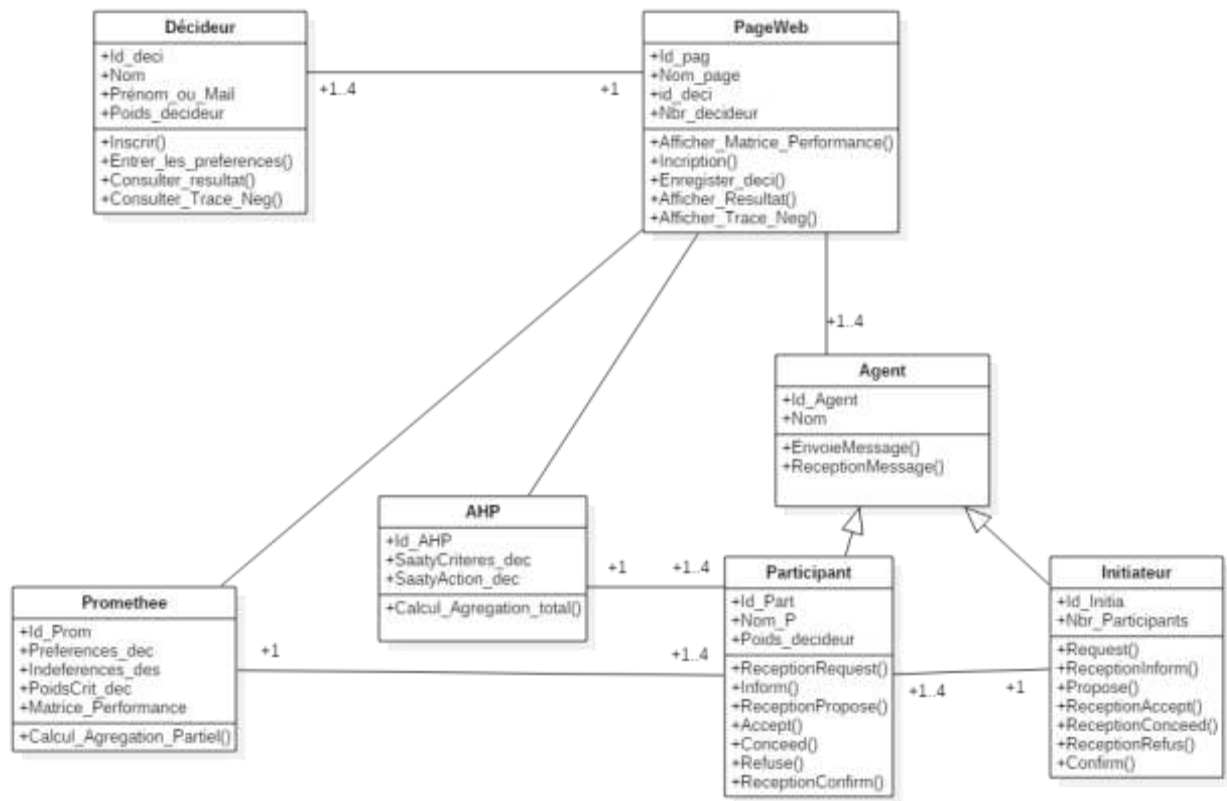

**Figure III.6** : Diagramme de classes.

Pour notre diagramme de classe, nous avons définit le dictionnaire de données afin d'expliquer les mots clés dans le diagramme.

- **Le diagramme de cas d'utilisation**

Le diagramme de cas d'utilisation (Use case) est l'un des diagrammes d'UML, il permet de recenser les grandes fonctionnalités d'un système.

Pour modéliser notre système, deux acteurs sont à considérer : les décideurs, et la page web (voir la Figure III.4).

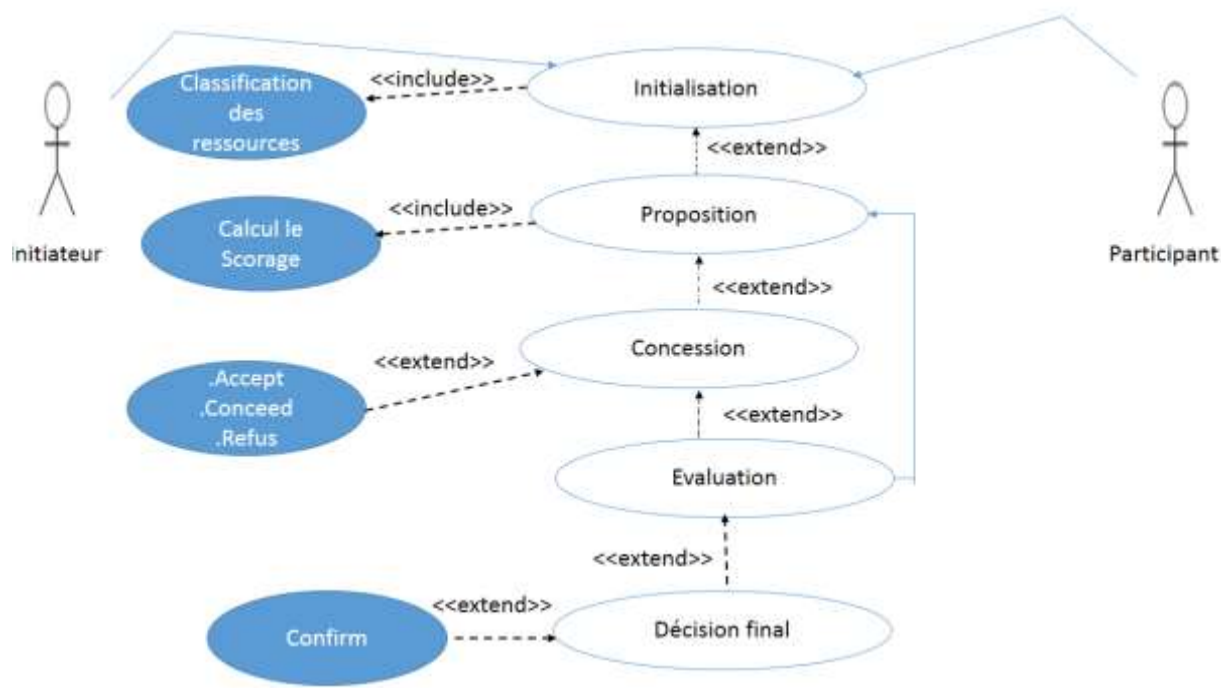

**Figure III.7** : Diagramme de cas d'utilisation du protocole proposé.

### 8.2.1. Description du diagramme de Use Case

Le tableau suivant montre (Tableau III.4) la description de notre diagramme de use case.

|  | **Description** |
|---|---|
| **Titre** | La négociation par le protocole de concession monotone. |
| **Acteurs Principaux** | L'Initiateur, les participants, la page Web. |
| **Prés-conditions** | <ul><li>Les agents participants et l'agent initiateur sont créés.</li><li>Les méthodes multicritères sont implémentées.</li><li>Les préférences des décideurs sont introduites.</li></ul> |
| **Post condition** | <ul><li>Nombre d'accept par les participants est supérieur ou égal au seuil de négociation.</li><li>Tous les agents participants ont la solution finale à la fin de la négociation.</li></ul> |
| **Résumé** | Ce use case permet à un ensemble des agents participants de négocier avec l'initiateur, pour trouver une solution finale de compromis. |

**Tableau III.4** : Tableau de description du diagramme de use case.

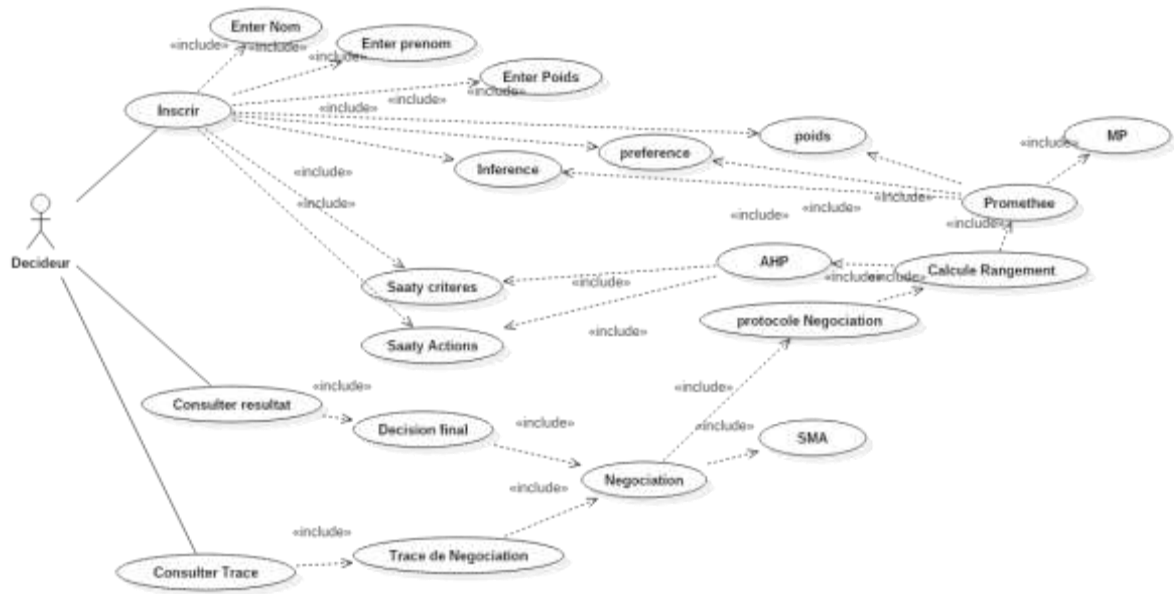

**Figure III.8** : Diagramme de cas d'utilisation de notre système d'aide à la décision de groupe.

- **Le diagramme de séquences**

Un diagramme de séquence indique l'interaction entre plusieurs partenaires de communication, également appelés lignes de vie. Une ligne de vie est représentée par un rectangle contenant son nom et un trait vertical en pointillés. Ce trait représente un axe temporel, orienté de haut en bas [UML 2, 09].

La figure III.5 représente le séquencement des primitives de notre protocole de négociation entre les participants et l'initiateur.

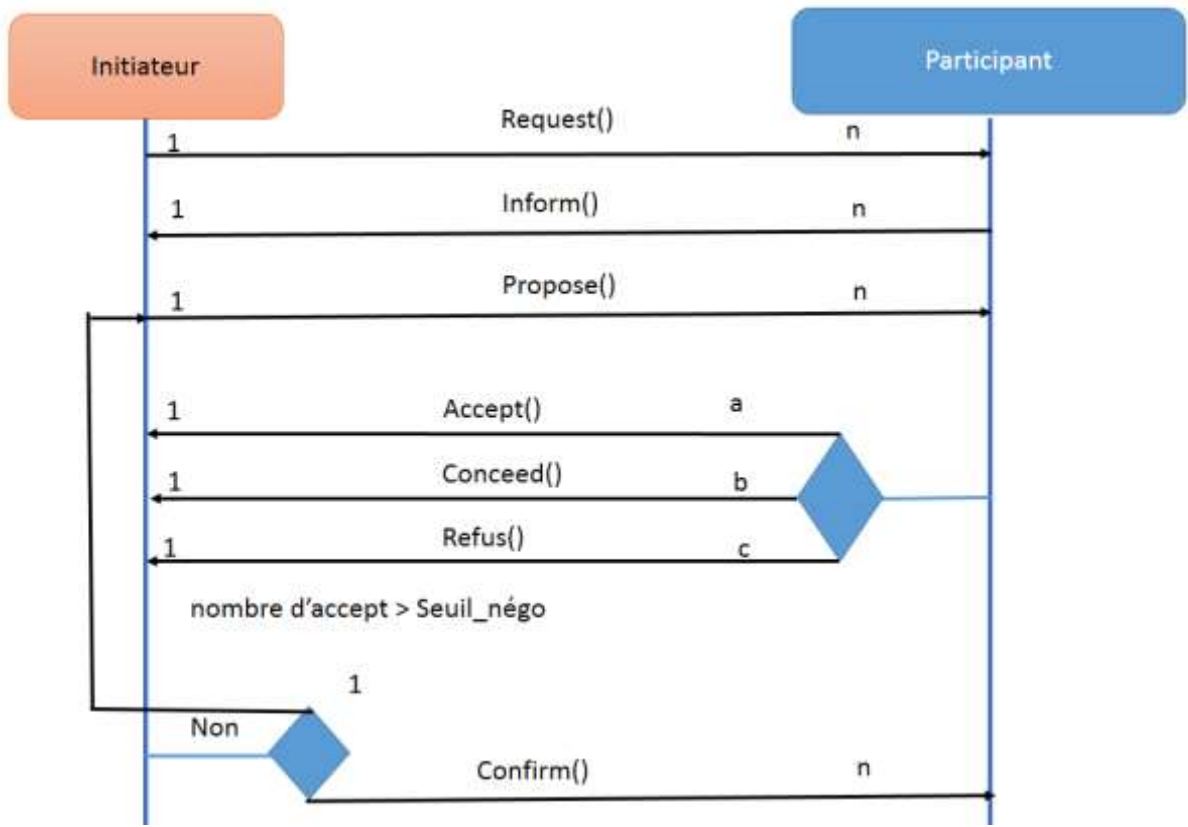

**Figure III.9** : Diagramme de séquences de notre protocole.

- **Le diagramme de collaboration**

Le diagramme de collaboration est un diagramme d'interaction, se concentrant sur les échanges des messages entre les objets avec plus de détails que le diagramme de séquences.

La figure suivante illustre le diagramme de collaboration du protocole de négociation :

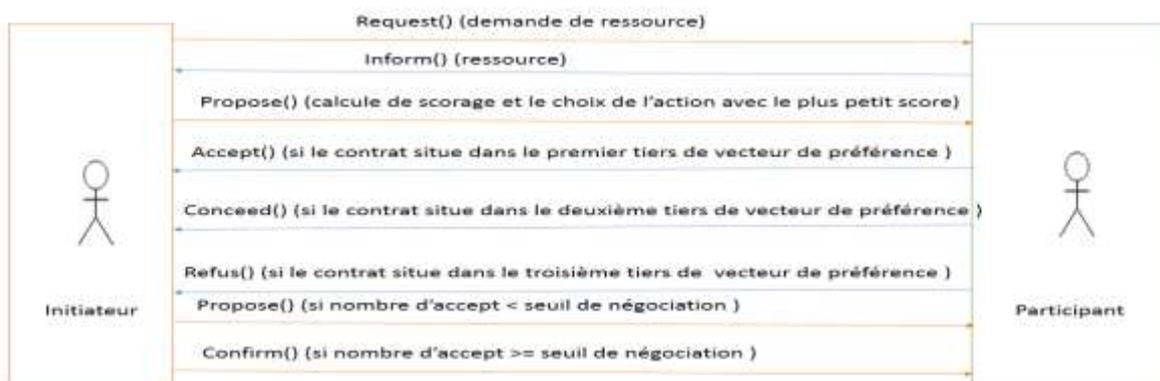

**Figure III.10 :** Diagramme de collaboration de notre protocole.

## 14.Conclusion

Nous avons présenté dans ce chapitre notre proposition de système d'aide à la décision de groupe et du protocole de négociation fondé sur les deux méthodes multicritères AHP, et Prométhée.

En premier lieu, nous avons posé notre problématique, puis nous avons proposé une architecture générale ainsi qu'une architecture fonctionnelle de notre système, ensuite nous avons décrit les méthodes multicritères AHP et Prométhée tout en citant leurs avantages et leurs inconvénients dans le cadre de notre problématique.

Nous avons, également, présenté les modèles correspondants sous forme des diagrammes UML à savoir : le diagramme de classes, le diagramme de séquences, le diagramme de cas d'utilisation, et le diagramme de collaboration.

Le chapitre suivant sera consacré à la mise en œuvre de notre application.

# Chapitre IV

# Mise en œuvre

1. **Introduction**

L'implémentation constitue la phase de réalisation proprement dite du système, c'est-à-dire l'écriture des programmes dans le langage de programmation approprié.

L'objectif de la phase d'implémentation est d'aboutir à un produit final, exploitable par les utilisateurs.

Dans ce chapitre, en premier lieu nous présentons les outils de développement tout en spécifiant l'environnement de travail et les langages de programmation utilisés.

En second lieu, nous décrirons notre protocole proposé basé sur les méthodes multicritères AHP (Processus d'analyse hiérarchique) et Prométhée II par la présentation d'un scénario illustrant une étude de cas réelle.

2. **Langages de programmation**

Pour réaliser notre système, nous avons exploité un langage de la programmation et un langage de balisage (HTML) décrits dans ce qui suit.

### 2.1. HTML

Le langage HTML 5 est le standard de W3C de langage HTML publié en 2009, le but de ce nouveau standard est de structurer et de donner du sens aux programmes, il permet également, couplé avec JavaScript, de créer des applications web, c'est notamment le cas d'applications comme Gmail ou Google document [THI, 12].

### 2.2. CSS

C'est le langage de mise en forme des sites web. Alors que le HTML permet d'écrire le contenu des pages web et de les structurer, le langage CSS s'occupe de la mise en forme et de la mise en page. C'est en CSS qui choisit notamment la couleur, la taille des menus et bien d'autres choses [MAT, 17].

### 2.3. JavaScript

Créé à l'origine par Netscape, ce langage de programmation est conçu pour traiter localement des événements provoqués par le lecteur (par exemple, lorsque le lecteur fait glisser la souris sur une zone de texte, cette dernière change de couleur). C'est un langage interprété, c'est-à-dire que le texte contenant le programme est analysé au fur et à mesure par l'interprète, partie intégrante du browser, qui va exécuter les instructions [DAN et al, 14].

### 2.4. PHP

PHP est un langage de programmation utilisé sur de nombreux serveurs pour prendre des décisions. C'est PHP qui décide du code HTML qui sera généré et envoyé au client à chaque fois [MAT, 17].

### 2.5. JAVA

Nous avons opté pour l'utilisation de JAVA, dans le développement de notre application, pour les raisons suivantes :

- ❖ Simplicité

Java est un langage simple, bien conçu pour un langage pleinement objet. Cela facilite la formation du personnel, et aussi la créativité, souvent bridée par la complexité des interfaces de programmation [Java, 08].

❖ Orienté objet

Java est un langage orienté objet. Il permet notamment l'encapsulation du code dans des classes, ce qui facilite l'implémentation d'applications analysées par la décomposition du problème en objets [Java, 08].

❖ Portable

Cette portabilité est réelle, et possible grâce à la notion de machine virtuelle (JVM : Java Virtual Machine) [Java, 08].

❖ Performant

Bien qu'interprété, ce langage est performant grâce aux optimisations du compilateur JIT (Just In Time) et à la simplicité de l'architecture [Java, 08].

## 3. Les outils de développement

Dans notre approche, notre choix a porté sur les outils de développement suivants :

### 3.1. WampServer

WampServer est une plate-forme de développement Web sous Windows pour des applications Web dynamiques à l'aide du serveur Apache2, du langage de scripts PHP et d'une base de données MySQL. Il possède également PHPMyAdmin pour gérer plus facilement vos bases de données [Net, 02].

### 3.2. IDE NetBeans

Entre les différents IDE (IntegratedDevelopmentEnvironment) en anglais ou EDI en français pour l'Environnement de Développement Intégré), nous avons choisi de travaillé avec Netbeans 8.02, cette partie est consacrée à l'implémentation des méthodesd'aide multicritères à la décision AHP et Prométhée.

NetBeans est un projet open source ayant un succès et une base d'utilisateur très large, une communauté en croissance constante, et près 100 partenaires mondiaux et des centaines de milliers d'utilisateur à travers le monde. Sun Microsystems a fondé le projet open source NetBeans en Juin 2000 et continue d'être le sponsor principal du projet [Net, 03].

### 3.3. JADE

Pour développer notre système, nous avons choisi la plateforme JADE(**J**ava **A**gent **DE**velopement framework) pour la mise en place de notre système multi agents.

❖ **Qu'est-ce qu'une plate-forme JADE ?**

JADE est une plate-forme multi-agents développée en Java par le CSELT (Groupe de recherche de Gruppo Telecom, Italie) qui a comme but la construction des systèmes multi-agents et la réalisation d'applications conformes à la norme FIPA. JADE comprend deux composantes de base : une plate-forme agents compatible FIPA et un paquet logiciel pour le développement des agents Java [BEH, 14].

❖ **L'Architecture de JADE**

La plateforme JADE est composée de conteneurs d'agents (agent container) qui peuvent être distribués sur le réseau (Figure IV.1). Les agents vivent dans les conteneurs qui sont des processus java qui fournissent tous les services nécessaires pour l'hébergement et l'exécution des agents. La plateforme doit disposer d'un conteneur principal qui représente le point de démarrage, et qui se charge de [BEH, 14] :

- Gérer la table des conteneurs qui contient la liste ainsi que les adresses de tous les conteneurs qui composent la plateforme.
- Gérer la GADT (Global Agent Description Table) qui contient la liste de tous les agents présents dans la plateforme, leur statut actuel, et leur localisation.
- Héberger l'AMS (Agent Management System) et DF (Directory Facilitator) qui fournissent la gestion des agents, et le service des pages blanches ainsi que le service des pages jaunes respectivement.

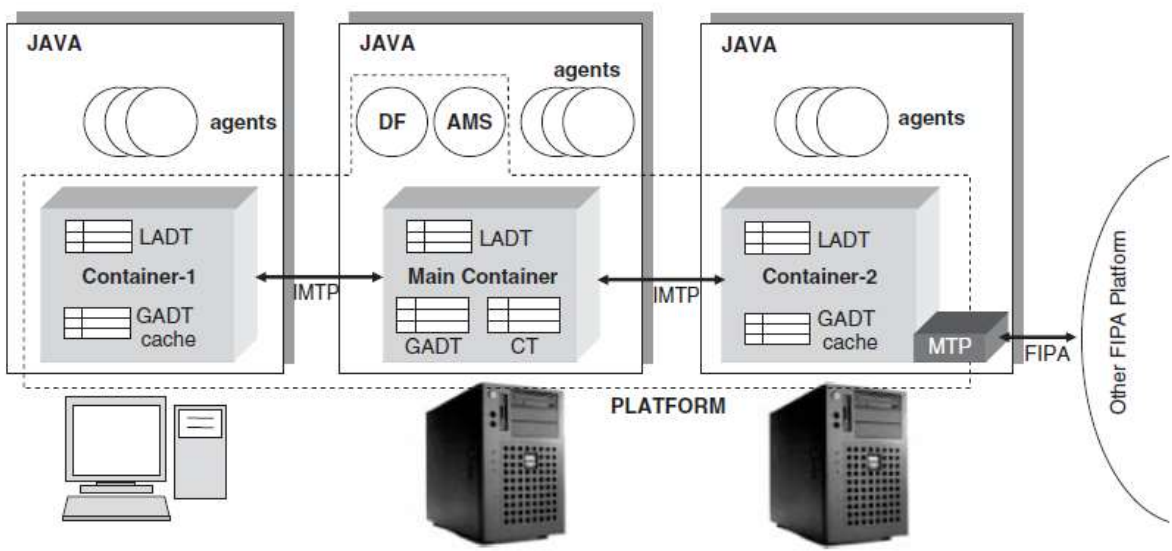

**Figure IV.**1 : Architecture de la plateforme JADE.

❖ **Outils de la plateforme JADE**

La plateforme jade est dotée d'un certain nombre d'outils tel que [BEH, 14]:

I. **Agent RMA (Remote Management Agent)**

Le RMA permet de contrôler le cycle de vie de la plate-forme et tous les agents la composant. Plusieurs RMA peuvent être lancés sur la même plate-forme du moment qu'ils ont des noms distincts. L'interface de l'agent RMA est illustrée par la figure IV.2

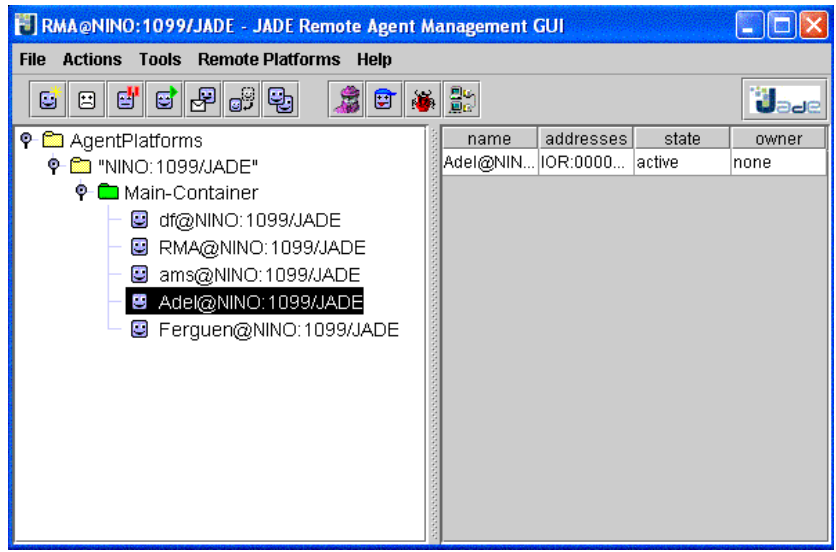

**Figure IV. 2:** L'interface de l'agent RMA.

II. **Agent Sniffer**

Quand un utilisateur décide d'épier un agent ou un groupe d'agents, il utilise un agent sniffer. Chaque message partant ou allant vers ce groupe est capté et affiché sur l'interface du sniffer. L'utilisateur peut voir et enregistrer tous les messages, pour éventuellement les analyser plus tard [BEH, 14]. L'interface de l'agent Sniffer est illustrée par la figure suivante :

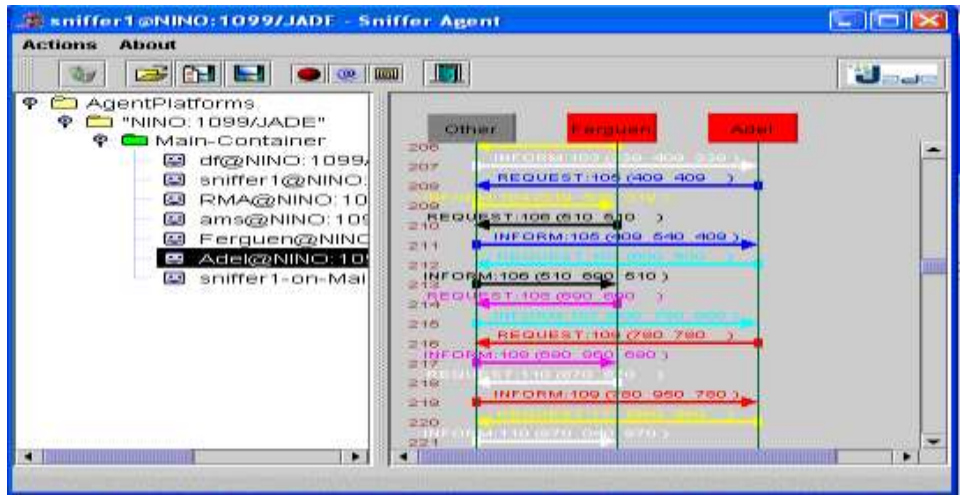

**Figure IV.3:** L'interface de l'agent Sniffer.

## 4. Etude de cas et résultats expérimentaux

Le protocole que nous avons développé peut être utilisé pour résoudre n'importe quel problème décisionnel de groupe où plusieurs critères conflictuels et plusieurs décideurs en conflit sont impliqués. Dans les sections suivantes, nous allons présenter le comportement de notre système pour la résolution d'une problématique décisionnelle réelle dans le domaine de l'aménagement du territoire.

### 4.1. Identification de problème décisionnel

La problématique traitée dans cette étude de cas est celle relative à la conception d'une carte d'adéquation du territoire pour l'habitat. En effet, la décision portera sur le choix du site le plus adéquat, parmi plusieurs pour la construction d'une habitation, ainsi nous disposons de 18 actions potentielles pouvant convenir pour cette construction [BEN, 15].

#### 4.1.1. Matrice de performance

Le tableau suivant représente la matrice des performances relative au problème posé :

| Actions | NUISANCES | BRUIT | IMPACTS | GEOTECHNIQ | EQUIPEMENT | ACCESSIBIL | CLIMAT |
|---|---|---|---|---|---|---|---|
| 729 | 1,00 | 0,99 | 2 | 6 | 1867 | 10 | 0,68 |
| 732 | 1,00 | 0,98 | 2 | 6 | 1957 | 10 | 0,71 |
| 737 | 1,00 | 0,97 | 2 | 6 | 2047 | 10 | 0,70 |
| 740 | 1,00 | 0,97 | 2 | 6 | 2147 | 10 | 0,69 |
| 743 | 1,00 | 0,93 | 2 | 6 | 2233 | 9 | 0,67 |
| 745 | 1,00 | 0,96 | 2 | 6 | 2185 | 12 | 0,84 |
| 748 | 1,00 | 0,67 | 2 | 6 | 2220 | 9 | 0,68 |
| 1030 | 1,00 | 0,15 | 4 | 6 | 1832 | 11 | 0,71 |
| 1033 | 0,99 | 0,55 | 4 | 6 | 1906 | 10 | 0,74 |
| 1038 | 0,98 | 0,27 | 4 | 6 | 2037 | 10 | 0,75 |
| 1045 | 1,00 | 0,96 | 4 | 6 | 2232 | 13 | 0,86 |
| 1046 | 1,00 | 0,69 | 4 | 6 | 2186 | 5 | 0,78 |
| 1233 | 1,00 | 0,62 | 6 | 3 | 1911 | 10 | 0,70 |
| 1236 | 1,00 | 1,00 | 6 | 3 | 2070 | 6 | 0,85 |
| 1239 | 1,00 | 1,00 | 6 | 3 | 2142 | 6 | 0,85 |
| 1321 | 1,00 | 0,98 | 6 | 6 | 1648 | 10 | 0,84 |
| 1324 | 1,00 | 0,98 | 6 | 6 | 1756 | 10 | 0,68 |
| 1326 | 1,00 | 0,98 | 6 | 6 | 1821 | 10 | 0,83 |

**Tableau IV.1 :** Matrice de performance.

### 4.1.2. Identification des décideurs

Pour cet exemple, nous identifions quatre décideurs impliqués dans la décision de groupe possédant chacun ses propres préférences exprimés par des paramètres subjectifs . Ces derniers sont spécifiques pour chaque méthode utilisée.

**1- Paramètres exprimés pour la méthode Prométhée II**

| *Paramètres Decideur1* | Nuisance | Bruit | Impact | Géotechnique | Equipement | Accessibilité | Climat |
|---|---|---|---|---|---|---|---|
| *Poids* | 7.51 | 13.63 | 13.63 | 13.63 | 17.2 | 17.2 | 17.2 |
| *Indifférence* | 0.3 | 0.3 | 0 | 55 | 5 | 0.3 | 0.3 |
| *Préférence* | 0.6 | 0.6 | 0 | 110 | 10 | 0.6 | 0.6 |

**Tableau IV.2** : Paramètres subjectifs exprimés par le décideur1

| *Paramètres Decideur2* | Nuisance | Bruit | Impact | Géotechnique | Equipement | Accessibilité | Climat |
|---|---|---|---|---|---|---|---|
| *Poids* | 4.51 | 7.08 | 17.31 | 18.63 | 18.93 | 17.52 | 15.27 |
| *Indifférence* | 0.35 | 0.35 | 0.3 | 5 | 4 | 0.5 | 0.35 |
| *Préférence* | 0.7 | 0.7 | 0.6 | 110 | 8 | 1 | 0.7 |

**Tableau IV.3** : Paramètres subjectifs exprimés par le décideur2

| *Paramètres Decideur3* | Nuisance | Bruit | Impact | Géotechnique | Equipement | Accessibilité | Climat |
|---|---|---|---|---|---|---|---|
| *Poids* | 6.15 | 19.57 | 13.79 | 13.79 | 13.79 | 16.45 | 16.45 |
| *Indifférence* | 0.2 | 0.2 | 0.1 | 30 | 2 | 0.15 | 0.2 |
| *Préférence* | 0.4 | 0.4 | 0.2 | 60 | 4 | 0.6 | 0.4 |

**Tableau IV.4**: Paramètres subjectifs exprimés par le décideur3

| *Paramètres Decideur4* | Nuisance | Bruit | Impact | Géotechnique | Equipement | Accessibilité | Climat |
|---|---|---|---|---|---|---|---|
| *Poids* | 17.38 | 29.4 | 6.16 | 6.16 | 6.16 | 17.38 | 17.38 |
| *Indifférence* | 0.25 | 0.3 | 0.15 | 45 | 3 | 0.25 | 0.25 |
| *préférence* | 0.5 | 0.6 | 0.3 | 90 | 6 | 0.5 | 0.5 |

**Tableau IV.5** : Paramètres subjectifs exprimés par le décideur4

2- **Paramètres exprimés pour la méthode AHP**

| *Critère\*Critère Décideur1* | Nuisance | Bruit | Impact | Géotechnique |
|---|---|---|---|---|
| *Nuisance* | 1 | 0.33 | 0.14 | 0.14 |
| *Bruit* | 3 | 1 | 0.33 | 5 |
| *Impact* | 7 | 3 | 1 | 3 |
| *Géotechnique* | 7 | 0.2 | 0.33 | 1 |

**Tableau IV.6** : Paramètres exprimés par le décideur1.

| *Critère : Nuisance Décideur1* | Action1 | Action2 | Action3 | Action4 |
|---|---|---|---|---|
| *Action1* | 1 | 7 | 7 | 0.33 |
| *Action2* | 0.14 | 1 | 7 | 5 |
| *Action3* | 0.14 | 0.14 | 1 | 0.33 |
| *Action4* | 3 | 0.2 | 3 | 1 |

**Tableau IV. 7**: Paramètres subjectifs decideur1 de critère Nuisance.

| *Critère : Bruit Décideur1* | Action1 | Action2 | Action3 | Action4 |
|---|---|---|---|---|
| *Action1* | 1 | 7 | 3 | 7 |
| *Action2* | 0.14 | 1 | 3 | 5 |
| *Action3* | 0.33 | 0.33 | 1 | 5 |
| *Action4* | 0.14 | 0.2 | 0.2 | 1 |

**Tableau IV.8** : Paramètres subjectifs decideur1 de critère Bruit

| *Critère :* Impact Décideur1 | Action1 | Action2 | Action3 | Action4 |
|---|---|---|---|---|
| *Action1* | 1 | 0.14 | 0.2 | 0.11 |
| *Action2* | 7 | 1 | 7 | 9 |
| *Action3* | 5 | 0.14 | 1 | 5 |
| *Action4* | 9 | 0.11 | 0.2 | 1 |

**Tableau IV.9** : Paramètres subjectifs decideur1 de critère Impact.

| *Critère :* Géotechnique Décideur1 | Action1 | Action2 | Action3 | Action4 |
|---|---|---|---|---|
| *Action1* | 1 | 5 | 1 | 3 |
| *Action2* | 0.2 | 1 | 7 | 0.33 |
| *Action3* | 1 | 0.14 | 1 | 0.11 |
| *Action4* | 0.33 | 3 | 9 | 1 |

**Tableau IV.10** : Paramètres subjectifs decideur1 de critère Géotechnique.

| *Critère*Critère* Décideur2s | Nuisance | Bruit | Impact | Géotechnique |
|---|---|---|---|---|
| *Nuisance* | 1 | 5 | 0.14 | 3 |
| *Bruit* | 0.2 | 1 | 0.11 | 5 |
| *Impact* | 7 | 7 | 1 | 0.14 |
| *Géotechnique* | 0.33 | 0.2 | 7 | 1 |

**Tableau IV.11**: Paramètres subjectifs exprimés par le décideur2.

| *Critère :* Nuisance Décideur2 | Action1 | Action2 | Action3 | Action4 |
|---|---|---|---|---|
| *Action1* | 1 | 0.2 | 0.11 | 9 |
| *Action2* | 5 | 1 | 0.33 | 5 |
| *Action3* | 9 | 3 | 1 | 9 |
| *Action4* | 0.11 | 0.2 | 0.11 | 1 |

**Tableau IV.12**: Paramètres subjectifs exprimés par le décideur2 pour le critère Nuisance.

| *Critère :* Bruit Décideur2 | Action1 | Action2 | Action3 | Action4 |
|---|---|---|---|---|
| *Action1* | 1 | 5 | 3 | 7 |
| *Action2* | 0.2 | 1 | 3 | 0.14 |
| *Action3* | 0.33 | 0.33 | 1 | 0.14 |
| *Action4* | 0.14 | 7 | 7 | 1 |

**Tableau IV.13**: Paramètres subjectifs exprimés par le décideur2 pour le critère Bruit.

| *Critère : Impact Décideur2* | Action1 | Action2 | Action3 | Action4 |
|---|---|---|---|---|
| *Action1* | 1 | 0.2 | 7 | 0.11 |
| *Action2* | 5 | 1 | 7 | 0.33 |
| *Action3* | 0.14 | 0.14 | 1 | 5 |
| *Action4* | 9 | 3 | 0.2 | 1 |

**Tableau IV.14**: Paramètres subjectifs exprimés par le décideur2 pour le critère Impact.

| *Critère : Géotechnique Décideur2* | Action1 | Action2 | Action3 | Action4 |
|---|---|---|---|---|
| *Action1* | 1 | 5 | 1 | 3 |
| *Action2* | 0.2 | 1 | 7 | 7 |
| *Action3* | 1 | 0.14 | 1 | 0.11 |
| *Action4* | 0.33 | 0.14 | 9 | 1 |

**Tableau IV.15**: Paramètres subjectifs exprimés par le décideur2 pour le critère Géotechnique.

| *Critère*Critère Décideur3* | Nuisance | Bruit | Impact | Géotechnique |
|---|---|---|---|---|
| *Nuisance* | 1 | 0.33 | 7 | 5 |
| *Bruit* | 3 | 1 | 3 | 5 |
| *Impact* | 0.14 | 0.33 | 1 | 9 |
| *Géotechnique* | 0.2 | 0.2 | 0.11 | 1 |

**Tableau IV.16** : Paramètres subjectifs exprimés par le décideur3.

| *Critère : Nuisance Décideur3* | Action1 | Action2 | Action3 | Action4 |
|---|---|---|---|---|
| *Action1* | 1 | 0.11 | 0.11 | 0.2 |
| *Action2* | 9 | 1 | 0.2 | 5 |
| *Action3* | 9 | 5 | 1 | 9 |
| *Action4* | 5 | 0.2 | 0.11 | 1 |

**Tableau IV.17**: Paramètres subjectifs exprimés par le décideur3 pour le critère Nuisance.

| *Critère : Bruit Décideur3* | Action1 | Action2 | Action3 | Action4 |
|---|---|---|---|---|
| *Action1* | 1 | 0.2 | 0.2 | 7 |
| *Action2* | 5 | 1 | 3 | 0.14 |
| *Action3* | 5 | 0.33 | 1 | 0.11 |
| *Action4* | 0.14 | 7 | 9 | 1 |

**Tableau IV.18**: Paramètres subjectifs exprimés par le décideur3 pour le critère Bruit.

| *Critère : Impact Décideur3* | Action1 | Action2 | Action3 | Action4 |
|---|---|---|---|---|
| *Action1* | 1 | 3 | 5 | 0.11 |
| *Action2* | 0.33 | 1 | 0.14 | 0.33 |
| *Action3* | 0.2 | 7 | 1 | 0.2 |
| *Action4* | 9 | 3 | 5 | 1 |

**Tableau IV.19**: Paramètres subjectifs exprimés par le décideur3 pour le critère Impact.

| *Critère : Géotechnique Décideur3* | Action1 | Action2 | Action3 | Action4 |
|---|---|---|---|---|
| *Action1* | 1 | 5 | 7 | 5 |
| *Action2* | 0.2 | 1 | 0.11 | 9 |
| *Action3* | 0.14 | 9 | 1 | 0.11 |
| *Action4* | 0.2 | 0.11 | 9 | 1 |

**Tableau IV.20**: Paramètre subjectifs exprimés par le décideur3 pour le critère Géotechnique.

| *Critère*Critère Décideur4* | Nuisance | Bruit | Impact | Géotechnique |
|---|---|---|---|---|
| *Nuisance* | 1 | 0.33 | 7 | 5 |
| *Bruit* | 3 | 1 | 7 | 5 |
| *Impact* | 0.14 | 0.14 | 1 | 0.2 |
| *Géotechnique* | 0.2 | 0.2 | 5 | 1 |

**Tableau IV.21**: Paramètres subjectifs exprimés par le décideur4.

| *Critère : Nuisance Décideur4* | Action1 | Action2 | Action3 | Action4 |
|---|---|---|---|---|
| *Action1* | 1 | 0.11 | 3 | 0.2 |
| *Action2* | 9 | 1 | 7 | 0.14 |
| *Action3* | 0.33 | 0.14 | 1 | 9 |
| *Action4* | 7 | 7 | 0.11 | 1 |

**Tableau IV.22**: Paramètres subjectifs exprimés par le décideur4 pour le critère Nuisance.

| *Critère : Bruit Décideur4* | Action1 | Action2 | Action3 | Action4 |
|---|---|---|---|---|
| *Action1* | 1 | 3 | 0.2 | 7 |
| *Action2* | 0.33 | 1 | 7 | 5 |
| *Action3* | 5 | 0.14 | 1 | 0.11 |
| *Action4* | 0.14 | 0.2 | 9 | 1 |

**Tableau IV.23**: Paramètres subjectifs exprimés par le décideur4 pour le critère Bruit.

| *Critère :  Impact*  *Décideur4* | Action1 | Action2 | Action3 | Action4 |
|---|---|---|---|---|
| *Action1* | 1 | 3 | 5 | 0.11 |
| *Action2* | 0.33 | 1 | 0.14 | 0.14 |
| *Action3* | 0.2 | 7 | 1 | 0.2 |
| *Action4* | 9 | 7 | 5 | 1 |

| *Critère :  Géotechnique*  *Décideur4* | Action1 | Action2 | Action3 | Action4 |
|---|---|---|---|---|
| *Action1* | 1 | 0.33 | 0.33 | 5 |
| *Action2* | 3 | 1 | 0.33 | 9 |
| *Action3* | 3 | 3 | 1 | 0.11 |
| *Action4* | 0.2 | 0.11 | 9 | 1 |

**Tableau IV.24**: Paramètres subjectifs exprimés par le décideur4 pour le critère

**Tableau IV.25**: Paramètres subjectifs exprimés par le décideur4 pour le critère Impact

## 5. Nos Fichiers des données

Nous avons utilisé plusieurs fichiers des données dont les structures sont décrites dans le Chapitre3.

### ❖ Pour la méthode AHP

Toutes les données nécessaires pour l'exécution de la méthode AHP sont stockées dans un fichier bien structuré, cette structure du fichier est montrée par la figure suivante :

```
mokhtar
omar
Politicien
Saaty_Critères 0.33 0.14 0.14 0.33 5 3
Saaty_Action1 7 7 0.33 7 5 0.33
Saaty_Action2 7 3 7 3 5 5
Saaty_Action3 0.14 0.2 0.11 7 9 5
Saaty_Action4 5 1 3 7 0.33 0.11
```

**Figure IV.4** : Fichier correspondant à la méthode AHP.

### ❖ Pour la méthode Prométhée II

Toutes les données nécessaires pour l'exécution de la méthode Prométhée II sont stockées dans un fichier bien structuré, cette structure du fichier est montrée par la figure suivante :

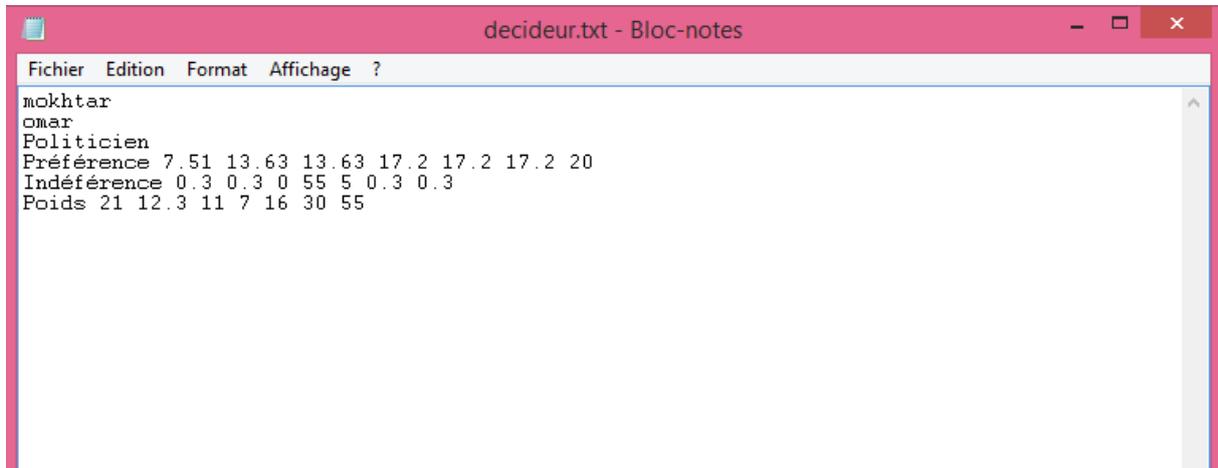

**Figure IV. 5:** Fichier correspondant à la méthode Prométhée.

## 6. Simulation de la négociation

Notre outil d'aide à la décision de groupe contient deux principaux modules : l'un est destiné pour la représentation des différents décideurs (interface web), et l'autre pour les traitements spécifiques.

La figure suivante illustre la page d'accueil de notre page web :

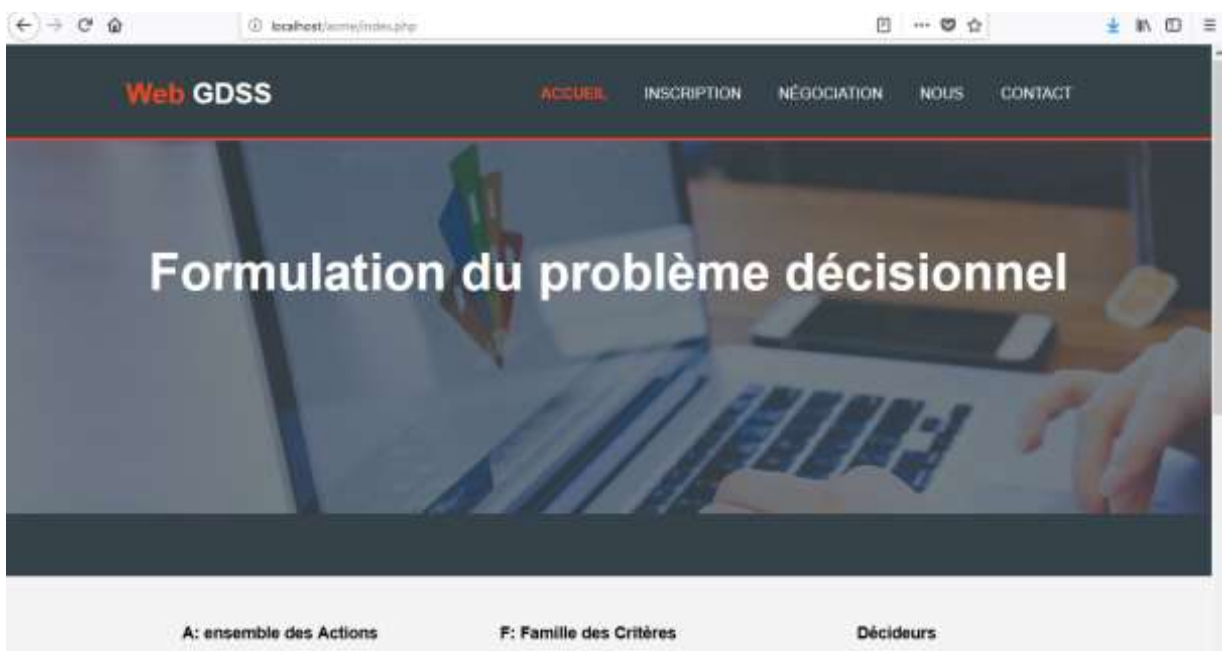

**Figure IV. 6:** Page Accueil de la page Web.

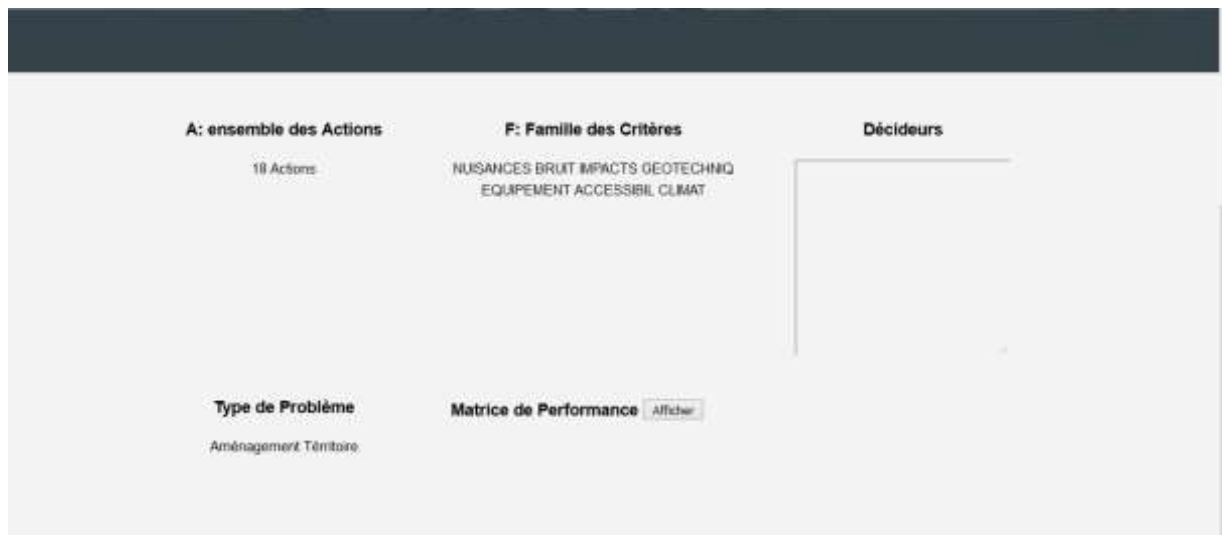

**Figure IV.7 :** Page Accueil de la page Web (formulation du problème décisionnel).

|        | NUISANCES | BRUIT | IMPACTS | GEOTECHNIQ | EQUIPEMENT | ACCESSIBIL | CLIMAT |
|--------|-----------|-------|---------|------------|------------|------------|--------|
| Alt1   | 1         | 0.99  | 2       | 6          | 1867       | 10         | 0.68   |
| Alt2   | 1         | 0.98  | 2       | 6          | 1957       | 10         | 0.71   |
| Alt3   | 1         | 0.97  | 2       | 6          | 2047       | 10         | 0.70   |
| Alt4   | 1         | 0.97  | 2       | 6          | 2147       | 10         | 0.69   |
| Alt5   | 1         | 0.93  | 2       | 6          | 2233       | 9          | 0.67   |
| Alt6   | 1         | 0.96  | 2       | 6          | 2185       | 12         | 0.84   |
| Alt7   | 1         | 0.97  | 2       | 6          | 2220       | 9          | 0.68   |
| Alt8   | 1         | 0.15  | 4       | 6          | 1832       | 11         | 0.71   |
| Alt9   | 0.99      | 0.55  | 4       | 6          | 1906       | 10         | 0.74   |
| Alt10  | 0.98      | 0.27  | 4       | 6          | 2037       | 10         | 0.75   |
| Alt11  | 1         | 0.96  | 4       | 6          | 2232       | 13         | 0.86   |
| Alt12  | 1         | 0.69  | 4       | 6          | 2186       | 5          | 0.78   |
| Alt13  | 1         | 0.62  | 6       | 3          | 1911       | 10         | 0.70   |
| Alt14  | 1         | 1     | 6       | 3          | 2070       | 6          | 0.85   |
| Alt15  | 1         | 1     | 6       | 3          | 2142       | 6          | 0.85   |
| Alt16  | 1         | 0.98  | 6       | 6          | 1648       | 10         | 0.84   |
| Alt17  | 1         | 0.98  | 6       | 6          | 1756       | 10         | 0.68   |
| Alt18  | 1         | 0.98  | 6       | 6          | 1821       | 10         | 0.83   |

**Figure IV. 8:** Structure de la Matrice des performances

**Figure IV. 9 :** Onglet d'inscription d'un décideur et introduction de ses préférences.

La partie mentionnée en rouge concerne l'introduction des paramètres subjectifs de la méthode Prométhée II.

**Figure IV.10 :** Onglet d'inscription d'un décideur et introduction de ses préférences (suite).

La partie mentionnée en rouge concerne l'introduction d es paramètres subjectifs de la méthode AHP.

Le bouton vert est utilisé pour valider l'inscription et la génération du fichier .Txt correspondant à ce décideur.

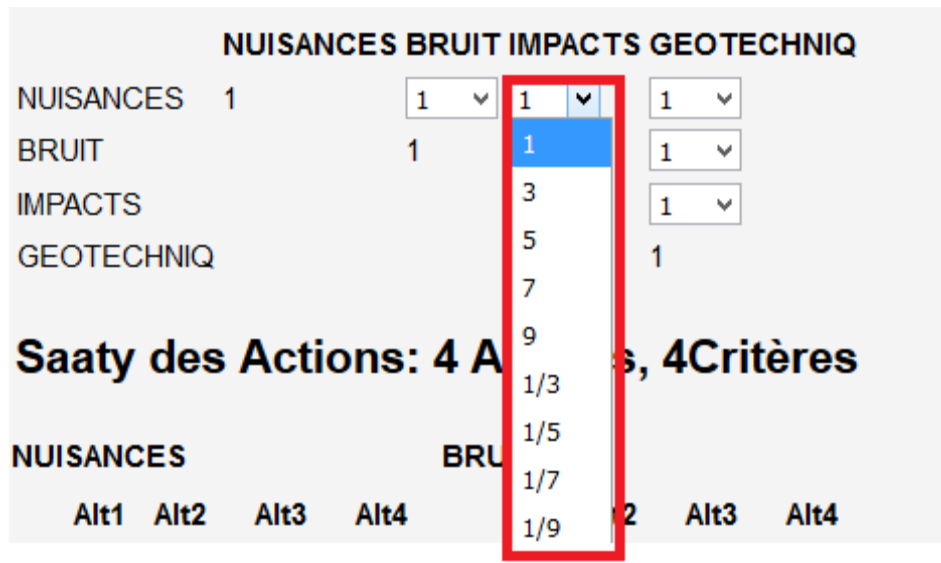

**Figure IV. 11**: L'échelle de Saaty proposée au décideur.

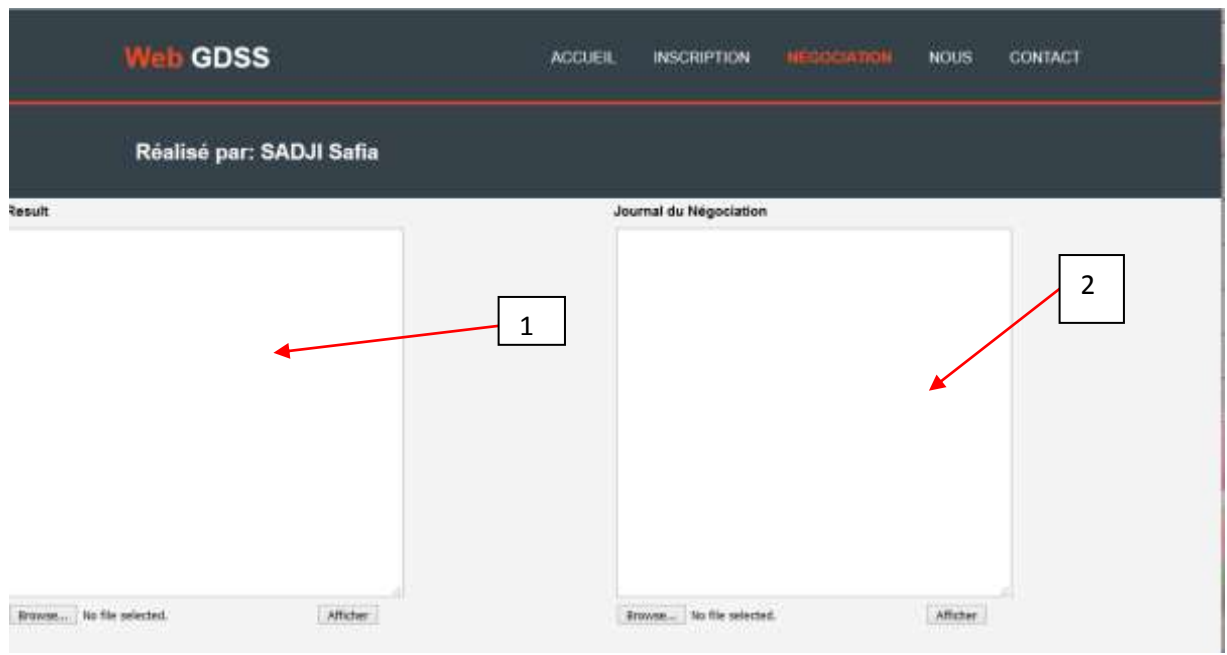

**Figure IV. 12 :** Onglet d'affichage du résultat aux décideurs

Dans cette figure (Figure IV. 12) :

- 1 correspond à un espace d'affichage du résultat final.
- 2 correspond à un espace d'affichage de l'historique de la négociation.

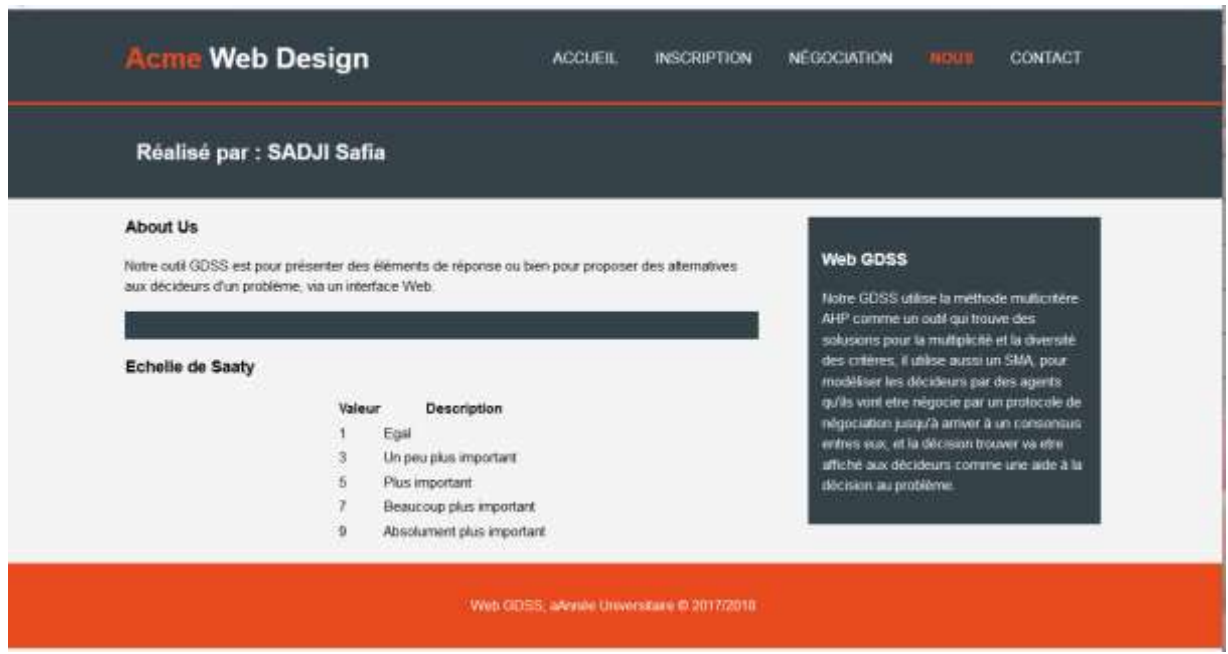

**Figure IV. 13 :** Onglet de l'aide.

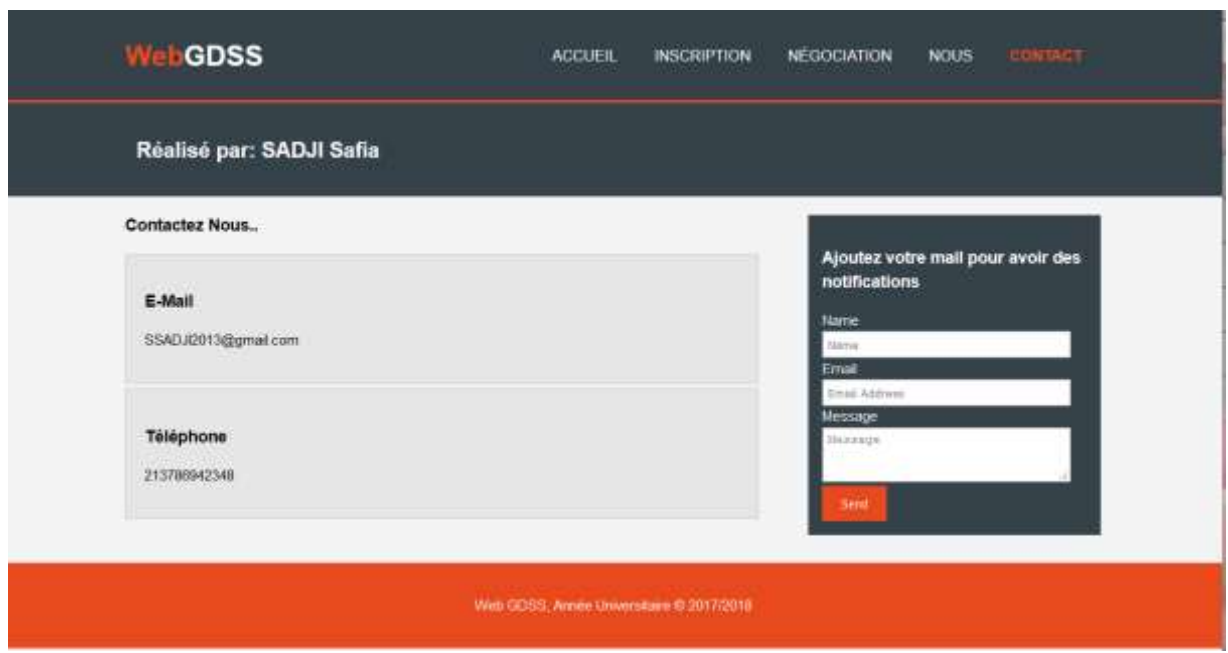

**Figure IV. 14 :** Onglet de contact.

❖ **Les interfaces de traitement :**

**Figure IV. 15 :** Représentation interface principale d'identification des décideurs (Prométhée).

En cliquant sur le bouton Matrice de performance, la fenêtre (Figure IV.15) :

En cliquant sur le bouton Ajouter, la fenêtre (Figure IV.17) s'affichera :

Les trois boutons « Promethe » « AHP » et « SMA » ne peuvent être actifs qu'après introduction de tous les paramètres subjectifs.

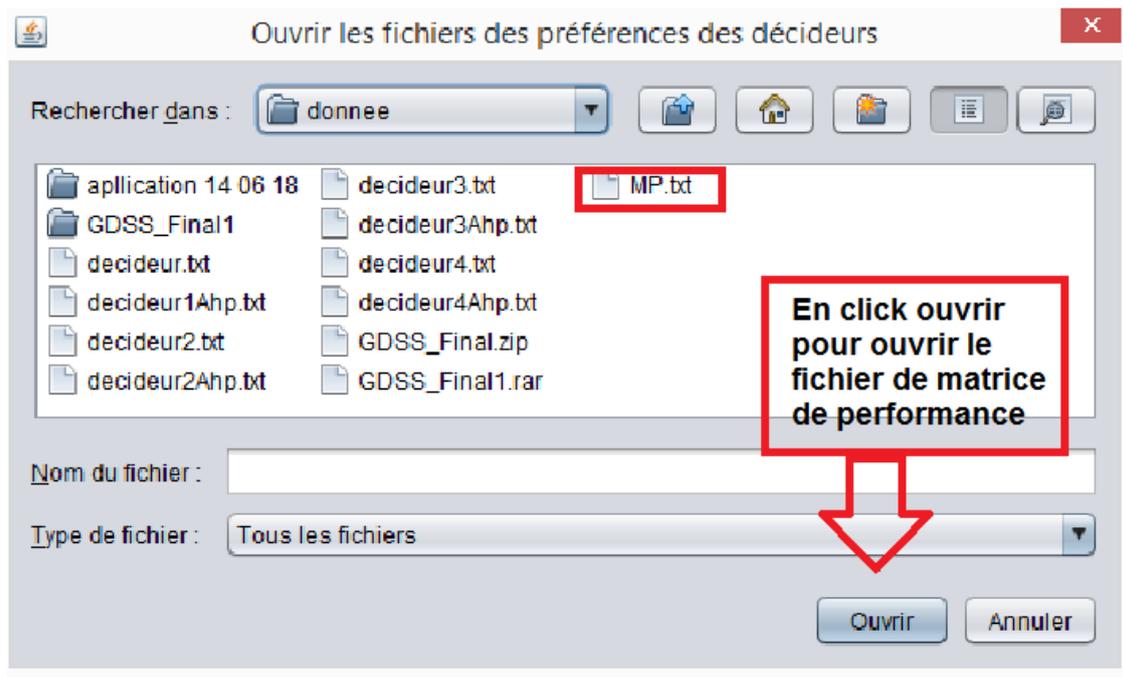

**Figure IV. 16:** Fenêtre pour ouvrir le fichier de matrice de performance.

**Figure IV. 17 :** Matrice de performance.

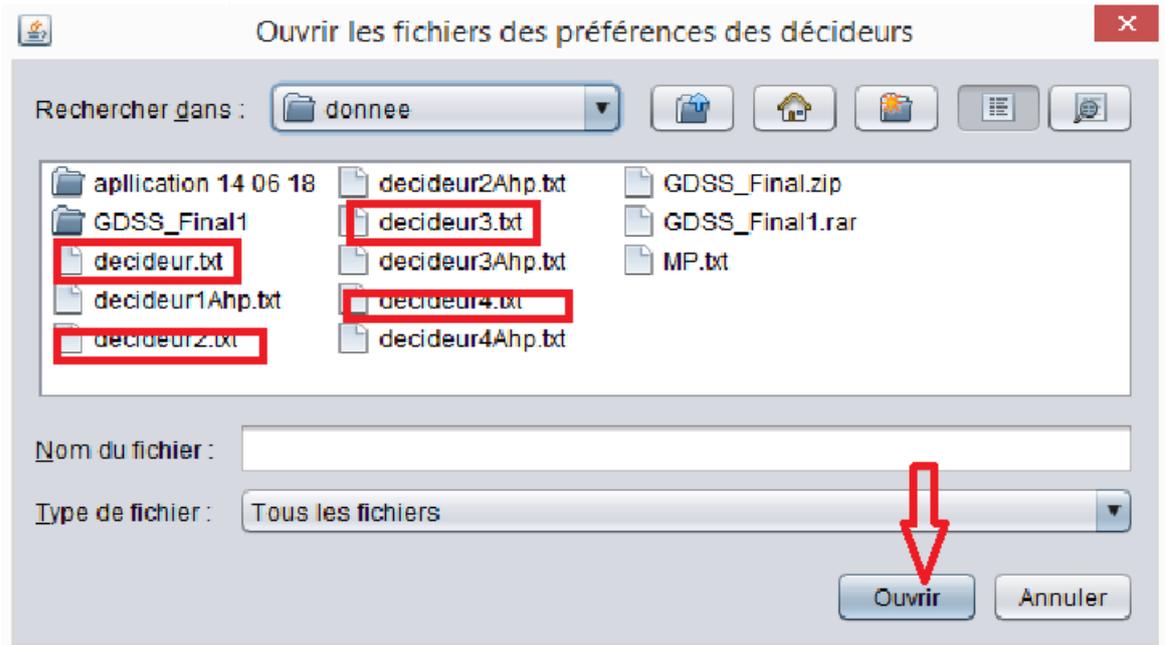

**Figure IV.18:** Fenêtre pour choisir le fichier .Txt d'un décideur utilisant –ProméthéeII-.

A chaque fois, on choisit un fichier .Txt parmi les fichiers sélectionnés à la figure IV.16 pour remplir les tableaux des préférences des décideurs comme la figure suivante montre :

**Figure IV.19** : Représentation interface principale.

A ce stade, les données de la méthode Prométhée II sont chargées, on clique, alors, sur le bouton « Prométhée » pour afficher le rangement des actions (Figure suivante)

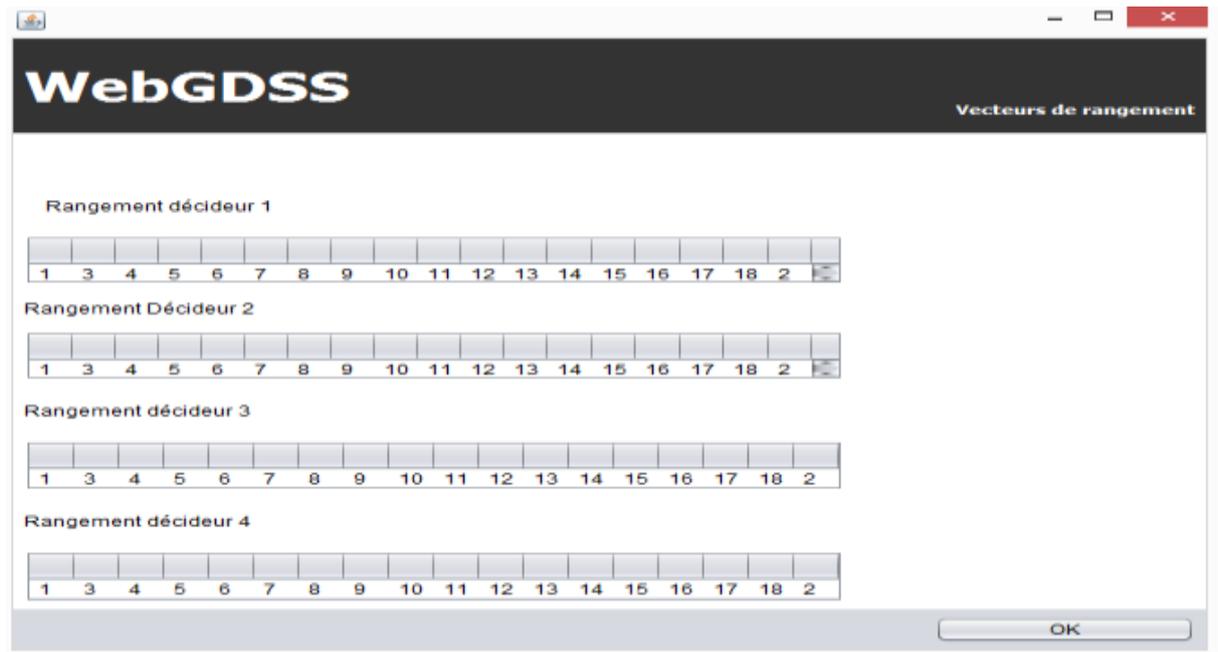

**Figure IV.20** : Représentation des rangements avec Prométhée Ill.

On revient sur l'interface principale pour calculer le rangement en utilisant la méthode AHP.

On click le bouton AHP, et la fenêtre suivante s'affichera :

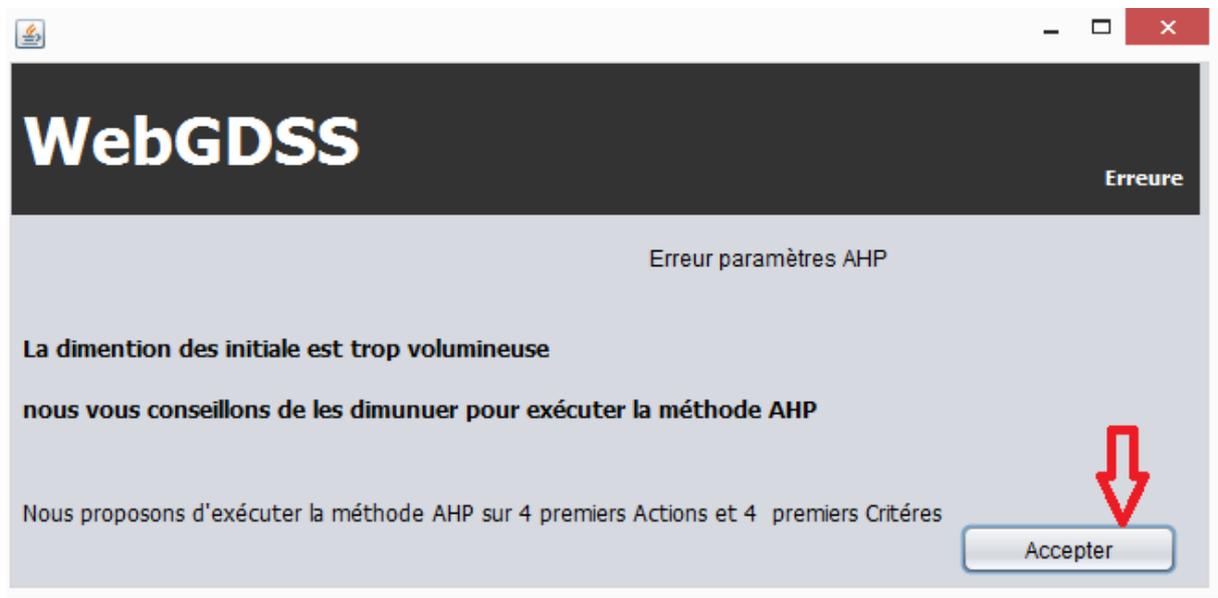

**Figure IV.21** : Fenêtre de confirmation pour pouvoir utiliser la méthode AHP.

On click sur le bouton Accepter pour confirmer d'utiliser la méthode AHP pour les quatre premiers actions et critères identifiés pour notre problème décisionnel, et la figure suivante s'affichera :

On click le bouton décideur i à chaque fois, et le résultat s'affichera comme suit :

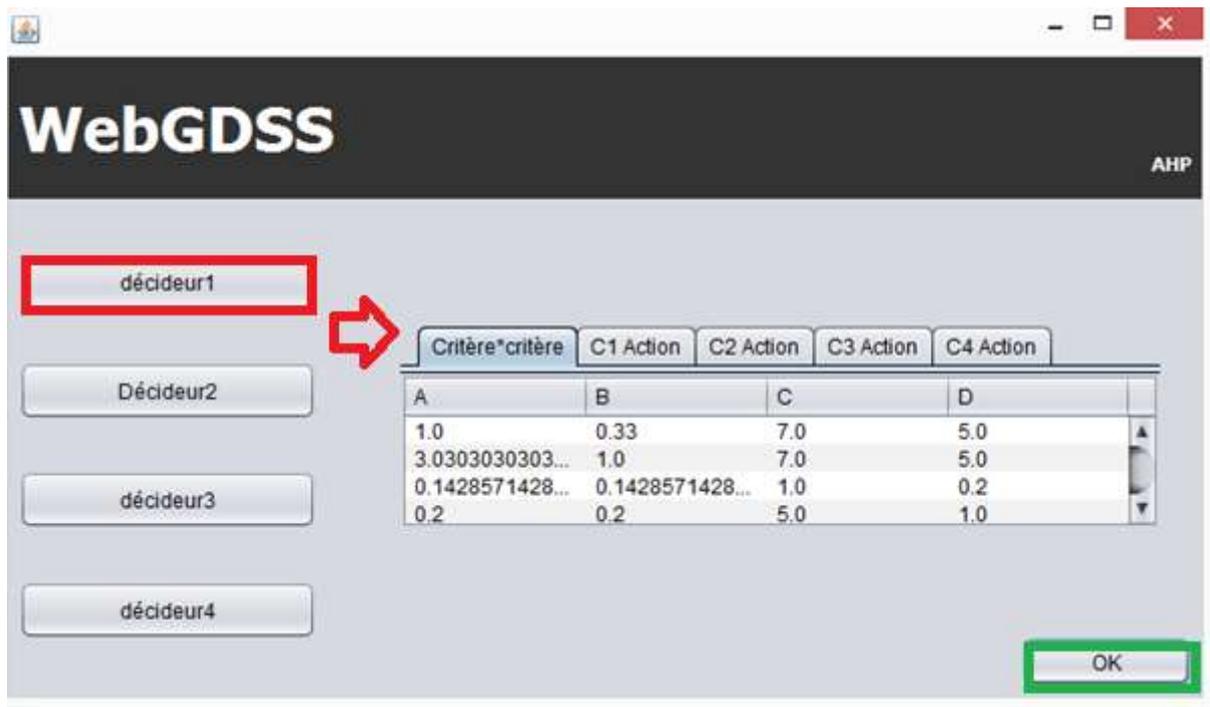

**Figure IV.22 :** Fenêtre pour choisir le fichier .Txt d'un décideur utilisant -AHP-.

**Figure IV.23 :** L'échelle de Saaty d'un décideur ( critère * critère).

**Figure IV.24 :** L'échelle de Saaty d'un décideur (action*action) / critère1.

**Figure IV.25 :** L'échelle de Saaty d'un décideur (action*action) / critère2.

**Figure IV.26 :** L'échelle de Saaty d'un décideur (action*action) / critère3

**Figure IV.27:** L'échelle de Saaty d'un décideur (action*action) / critère4

On click sur le bouton ok de cette interface (figure IV.23) et la fenêtre suivante s'affichera :

**Chapitre IV : Mise en œuvre**

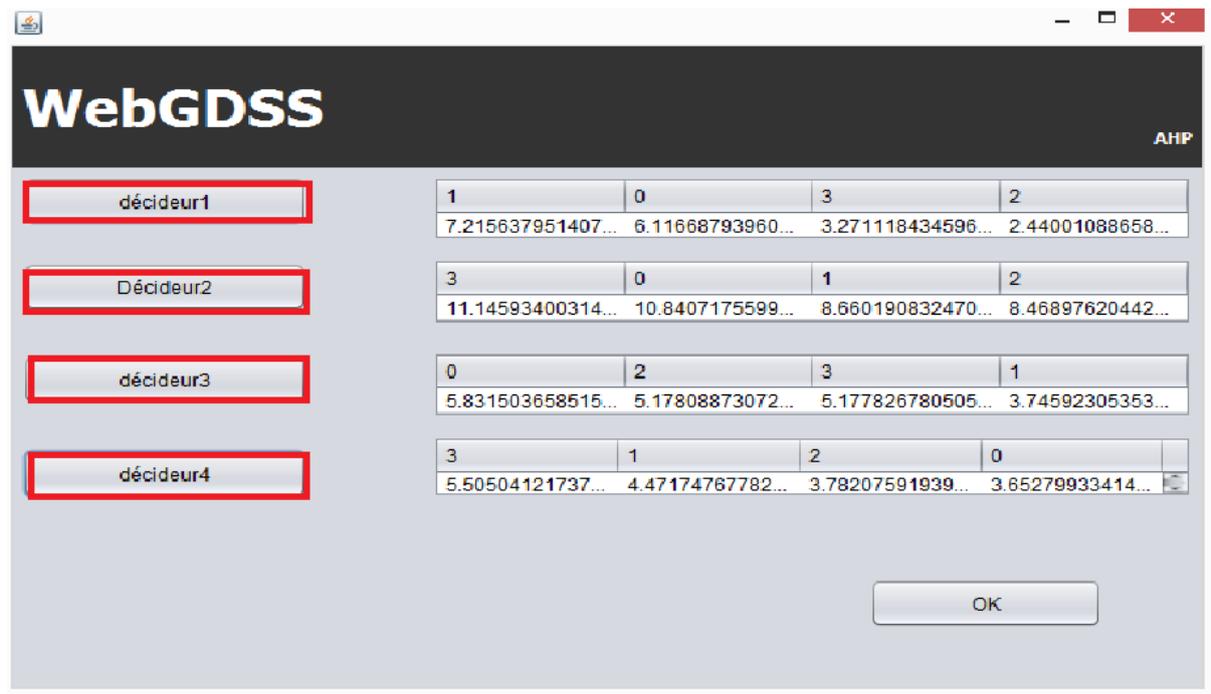

**Figure IV.28 :** Les rangements des actions obtenus avec AHP.

Afin de lancer la négociation, on click sur le bouton SMA de l'interface principale de Prométhée.

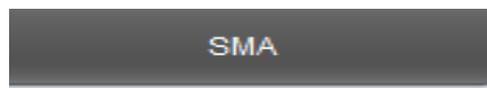

**Figure IV.29**: Le bouton SMA pour lancer la négociation.

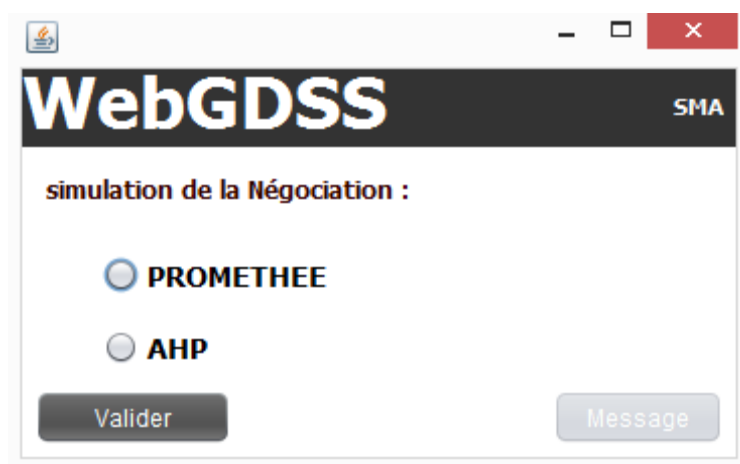

**Figure IV.30**: Fenêtre de choix entre AHP ou ProméthéeII pour la négociation.

Cette fenêtre est destinée pour lancer la négociation soit par l'utilisation de la méthode AHP soit par Prométhée II.

Si on ne choisit, ni AHP ni Prométhée, la fenêtre suivante s'affichera :

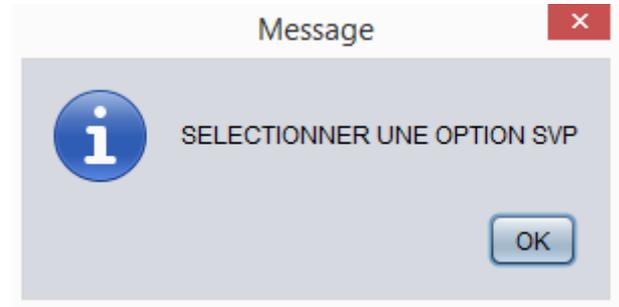

**Figure IV.31 :** Message d'information.

Nous proposons deux manières pour effectuer la négociation :

- ❖ Cas du choix de la méthode AHP :

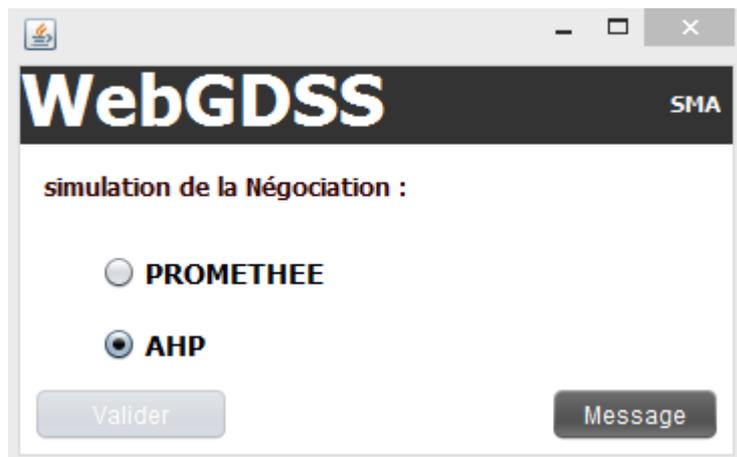

**Figure IV.32 :** Choix d'utilisation la méthode AHP.

Les messages échangés entre l'agent initiateur et les agents participants, au cours de la négociation, sont visualisés en utilisant l'agent Sniffer de JADE.

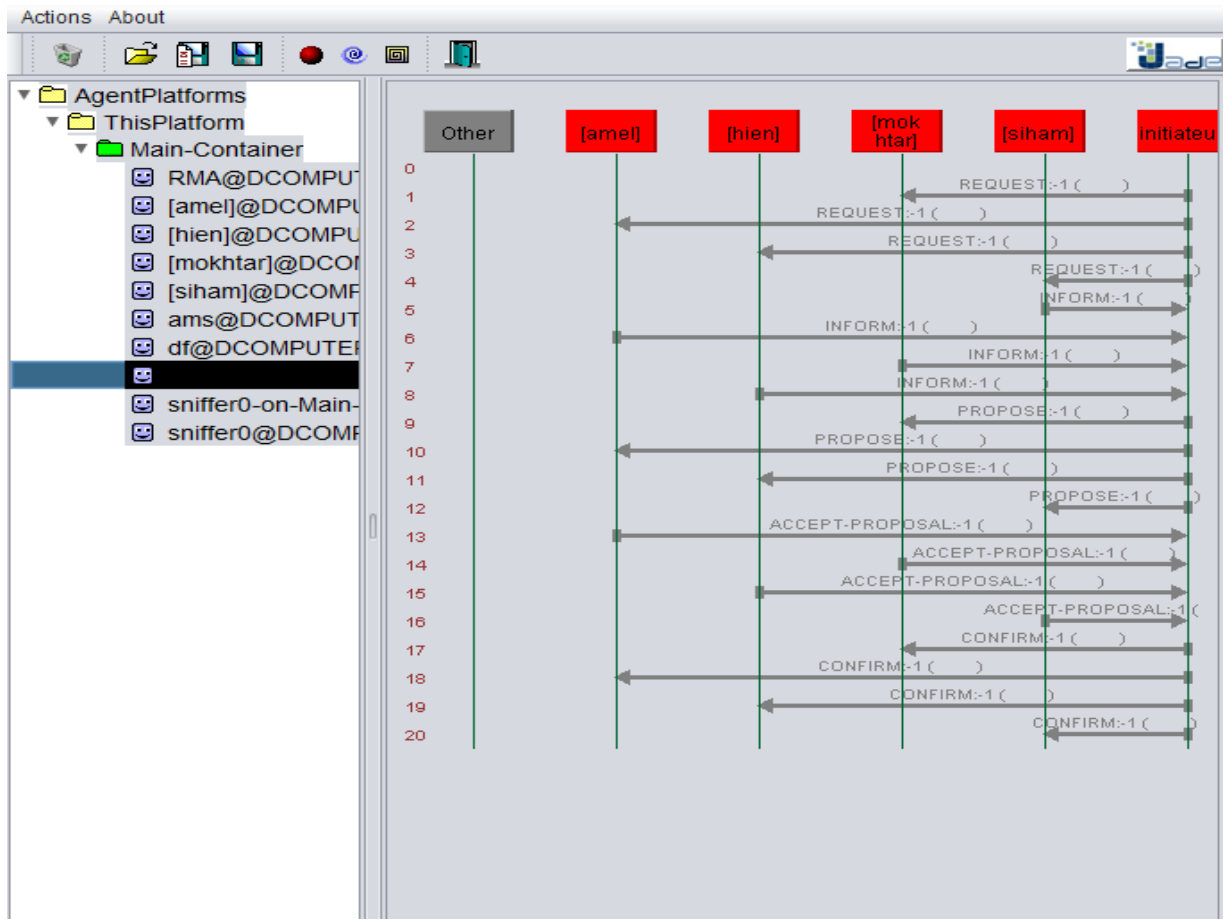

**Figure IV.33 :** Agent sniffer –AHP.

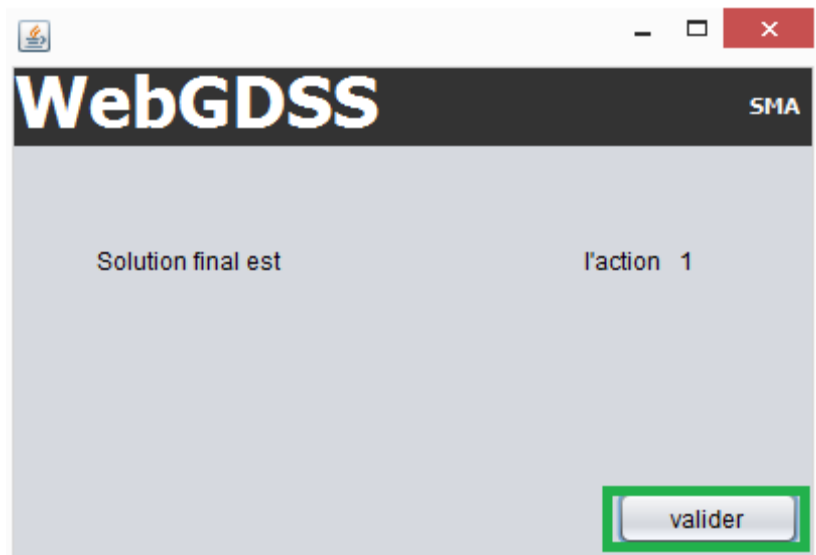

**Figure IV.34 :** Affichage de la solution final AHP.

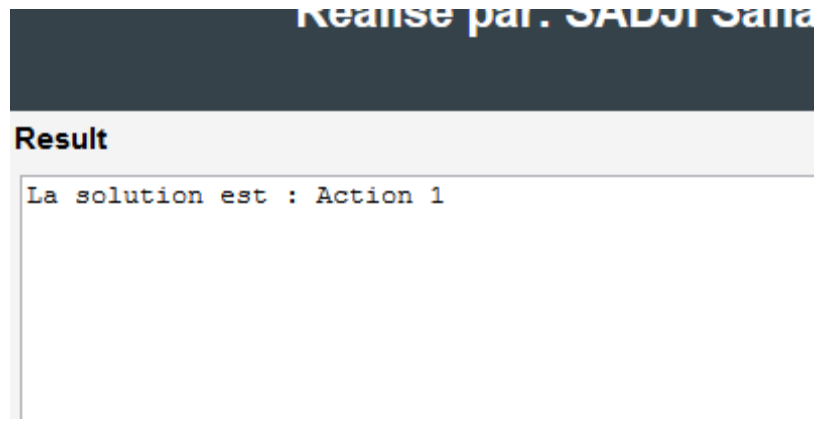

**Figure IV.35 :** Affichage de la solution final AHP aux décideurs

On clique le bouton valider, pour générer deux fichiers .Txt, un pour le résultat trouvé, et l'autre pour le trace de négociation.

Ces deux fichiers vont être affiché au l'interface négociation des décideurs.

- ❖ Cas de choix de la méthode Prométhée :

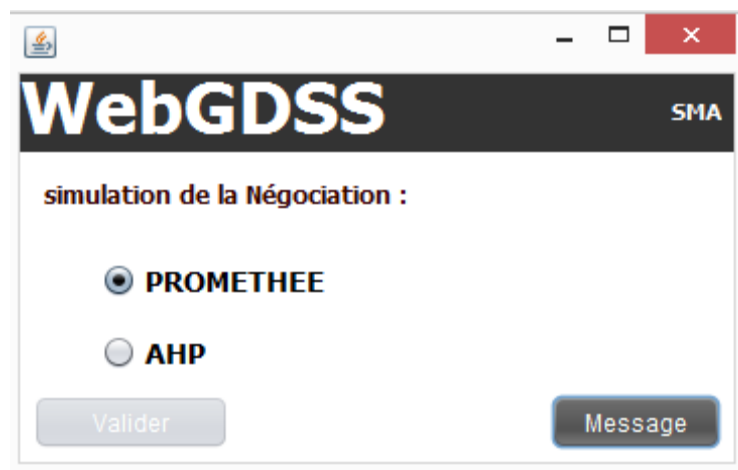

**Figure IV.36:** Choix d'utilisation la méthode Prométhée II

Les messages échangés entre l'agent initiateur et les agents participants, au cours de la négociation, sont visualisés en utilisant l'agent Sniffer de JADE.

**Chapitre IV : Mise en œuvre**

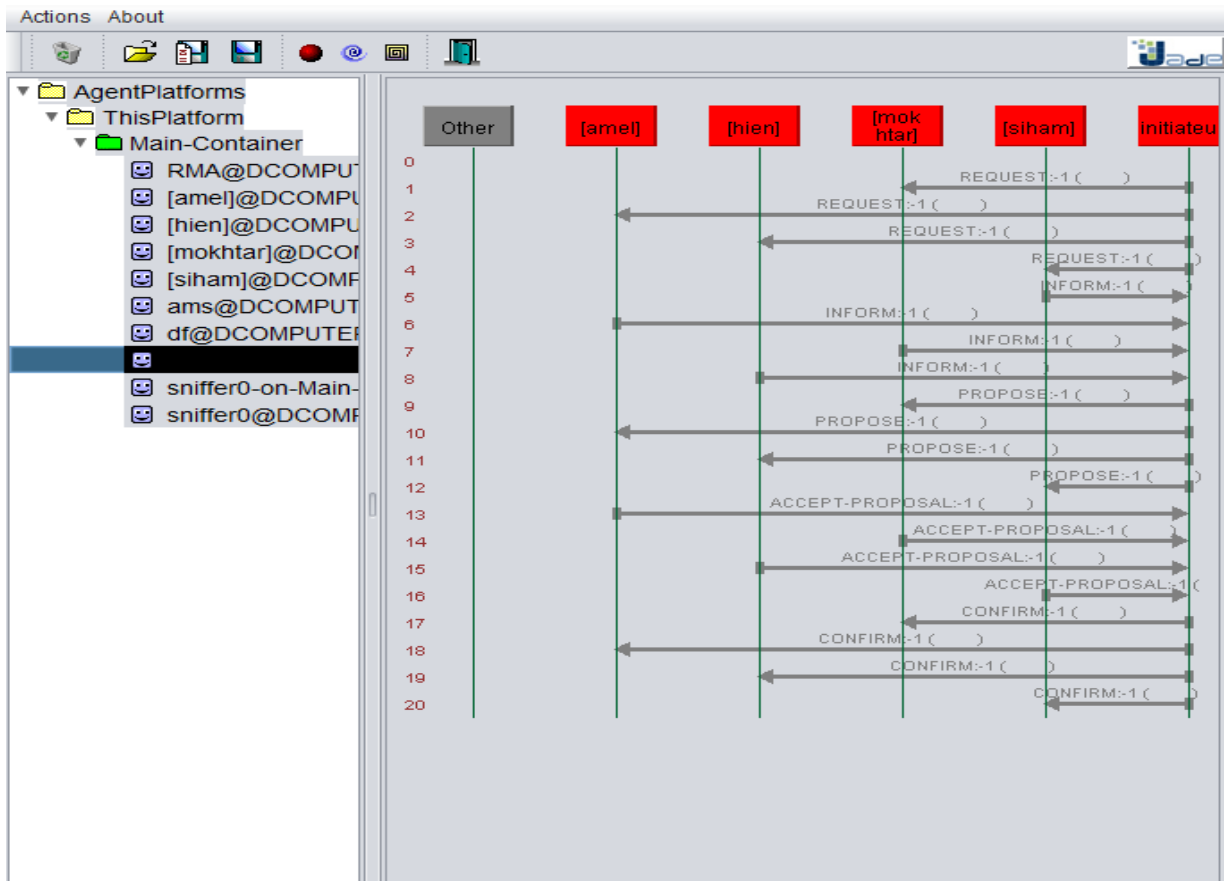

**Figure IV.37 :** Agent Sniffer –ProméthéeII

Lorsque l'initiateur émet un message « confirm » à tous les décideurs, la fenêtre suivanta s'affichera avec la solution trouvé :

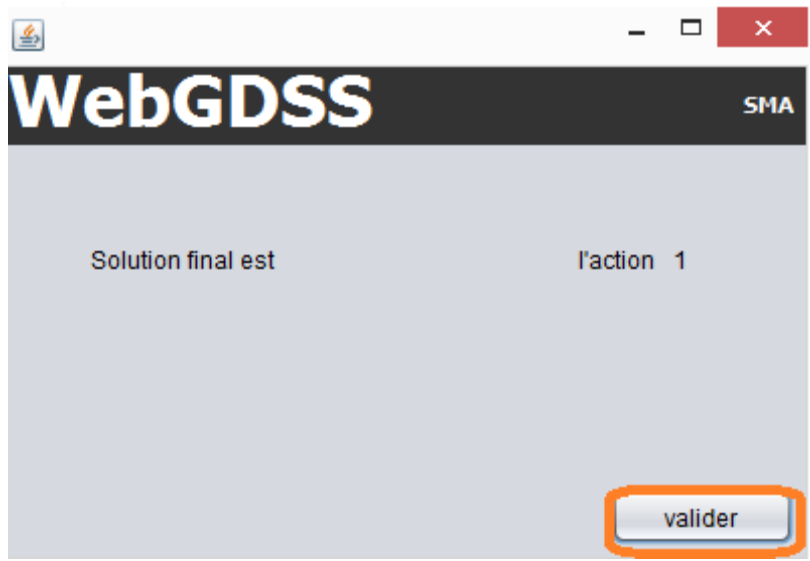

**Figure IV.38 :** Affichage de la solution final Prométhée

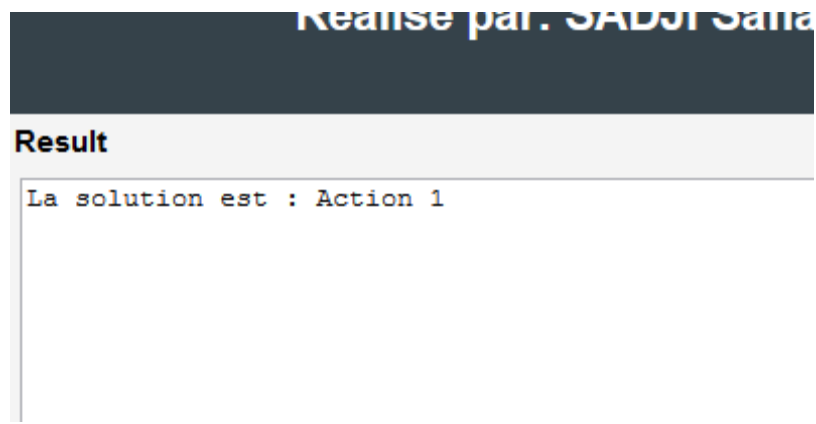

**Figure IV.39 :** Affichage de la solution final Prométhée

## 7. Conclusion

Dans ce chapitre, nous avons présenté la maquette informatique pour la mise en place de notre système d'aide à la décision de groupe GDSS. Ce dernier dispose d'une page web pour aider les décideurs à s'identifier et à exprimer convenablement leurs préférences tout en mettant à leur usage un protocole de négociation basé sur la médiation, la concession et l'agrégation multicritères d'aide à la décision.

# Conclusion Générale et Perspectives

Le thème abordé dans ce projet de fin d'étude est pluridisciplinaire et s'articule autour de plusieurs domaines d'intérêt à savoir : l'aide à la décision de groupe, l'analyse multicritères et la négociation dans les systèmes multi-agents. Les problèmes décisionnels traités, dans la présente étude, sont :

- A caractère multicritères nécessitant la définition de plusieurs critères conflictuels ;
- A caractères multidécideus : plusieurs acteurs sont impliqués dans la prise de décision, chacun a sa propre perception du problème, exprime ses propres préférences et défend ses propres objectifs ;
- A caractère distribué : les décideurs sont géographiquement dispersés.

Les modèles basés agents constituent une option innovante et prometteuse pour le développement d'outils d'aide à la décision multidécideurs. Ce paradigme permet de représenter la multiplicité des acteurs, leurs diversités, leur comportement ainsi que leur interactions.

A ce titre, la corrélation d'acteurs provenant de domaines différents souvent opposés disposant de leurs propres objectifs, intérêts et préférences nous a conduits alors à mettre en place une stratégie de négociation permettant de trouver un accord acceptable pour un groupe de décideurs.

## Contribution

Afin d'atteindre nos objectifs, nous nous sommes focalisés sur la définition d'un protocole de négociation basé sur la médiation et la concession, mettant en scène un agent initiateur responsable du bon déroulement de la négociation et un ensemble d'agents participants représentant les différents acteurs concernés par la décision.

*Conclusion Générale et Perspectives*

La première partie du mémoire constitue une synthèse de l'état de l'art composée de deux chapitres. Le premier chapitre a été consacré à la présentation des principaux concepts liés à l'aide à la décision de groupe. Le deuxième chapitre introduit, en détails, les différents concepts liés à la négociation dans les systèmes multi agents.

La deuxième partie étale le protocole proposé en deux chapitres. Le premier chapitre aborde les détails de la solution proposée. Le deuxième chapitre illustre les expérimentations effectuées et les résultats obtenus dans le but de mettre en valeur notre contribution.

Le point fort de notre approche est au niveau d'utilisation des méthodes d'analyse multicritères dans le processus de négociation. En effet, nous avons exploité deux méthodes différentes :

1. La méthode multicritères d'agrégation totale : AHP.
2. La méthode multicritères d'agrégation partielle : PROMETHEE II.

La solution proposée proposé permet de :

- Aborder l'aspect multicritères du problème décisionnel considéré en intégrant des méthodes d'aide multicritères à la décision;
- Représenter la multiplicité et la diversité des acteurs grâce aux capacités des SMA ;
- Prendre une décision collective en se basant sur une stratégie de négociation en proposant un protocole déterministe par envoi de messages;
- Offrir une Interface Web conviviale afin de faciliter l'interaction entre les décideurs

La solution proposée a été mise en œuvre en exploitant différents outils dont nous citons:

1. WampServer avec les langages PHP, JavaScript, HTML, et CSS pour le style.
2. IDE (Environnement de Développent Intégré) NetBeans, et avec langage java. l'environnement
3. JADE pour le système multi agents et la négociation.

## Perspectives

Nous terminons cette conclusion en évoquant quelques perspectives de recherche :

- Enrichir le système d'aide à la décision proposé en implémentant d'autres stratégies de négociation pour les agents participants et initiateurs.
- Exploiter d'autres plateformes de systèmes multi agents telle que JAVA Act et MADEKIT.
- Intégrer un module SIG afin de traiter des problématiques décisionnelles à caractère spatial.
- Mettre en œuvre une application client/serveur dédiée au système d'aide à la décision de groupe.
- Faire une étude comparative entre les résultats de plusieurs méthodes multicritères.

# Bibliographie

## Les Thèse et les mémoires

# Web graphie

|  | Lien | Date |
|---|---|---|
| **[Net, 01]** | **https://fr.wikipedia.org/wiki/StarUML** | 06/04/2018 |
| **[Net, 02]** | http://www.wampserver.com/ | 05/06/2018 |
| **[Net, 03]** | **https://netbeans.org/index_fr.html** , site officiel netbeans français. | 07/06/2018 |

## 1. Introduction

Avant l'apparition des méthodes multicritères, les problèmes de décisions se ramenaient le plus souvent à l'optimisation d'une fonction économique, cette approche vit le mérite de déboucher sur des problèmes mathématiques bien posés mais qui n'étaient pas toujours représentatifs de la réalité, car la comparaison de plusieurs actions possibles se fait rarement suivant un seul critère et les préférences sur un critère sont dans des cas difficilement modélisables par une fonction [HAM, 16].

Cette annexe constitue une présentation de l'analyse multicritères et les problèmes décisionnels ainsi que les méthodes utilisées pour arriver à la décision de compromis.

## 2. L'Analyse multicritères

D'après **Vincke**, l'analyse multicritère est : « *une approche constructivistes visant à fournir des outils permettant à progresser dans la résolution d'un problème ou plusieurs points de vue, souvent contradictoires, doivent être pris en compte* » [MAD, 11].

L'analyse multicritères repose sur un ensemble de procédures permettant de détailler un problème décisionnel portant sur des situations complexes. Dans l'analyse multicritères, on cherche un domaine de résolution pouvant tenir compte de l'ensemble des critères susceptibles d'influencer la décision. Le critère se définit comme un facteur à prendre en considération pour évaluer un scénario donné ou pour apprécier une occasion d'action [DOL,12]

## 3. Typologie des problèmes décisionnels

Les quatre problématiques décisionnelles de base sont présentées dans le tableau suivant [MAD, 11] :

| Problématique | Objectif | Résultat |
|---|---|---|
| P α | Eclairer la décision par le choix d'un sous-ensemble aussi restreint que possible en vue d'un choix finale d'une seul action, ce sous-ensemble contenant des « meilleurs » actions (optimaux) ou, à défaut, des actions « satisfaisantes ». | Un choix ou une procédure de sélection. |
| P β | Eclairer la décision par un tri résultant une affectation de chaque action à une catégorie étant définies a priori en fonction de normes ayant trait à la suite à donner aux actions qu'elles sont destinées à recevoir. | Un tri ou une procédure d'affectation. |
| P γ | Eclairer la décision par un rangement obtenu en regroupant tous ou partie (les « plus satisfaisantes ») des actions en en classes d'équivalence, ces classe étant ordonnées, de façon complète ou partielle, conformément aux préférences. | Un rangement ou une procédure de classement. |
| P Δ | Eclairer la décision par une description, dans un langage approprié, des actions et de leurs conséquences. | Une description ou une procédure cognitive. |

**Tableau 1 :** Les quatre problématiques décisionnelles de référence.

## 4. Terminologie associée à l'analyse multicritères

En l'analyse multicritère d'aide à la décision, on trouve plusieurs termes et des concepts de base, on présente ici le plus important :

### 4.1. Action

Selon B.ROY une action est défie comme étant *« la représentation d'une éventuelle contribution à la décision globale, susceptible, eu égard à l'état d'avancement du processus de décision, d'être envisagée d'une façon autonome et de servir de point d'application à l'aide à la décision »*[ALN, 16].

### 4.2. Critère

Un critère est une expression conçue pour évaluer et comparer des actions potentielles selon un point de vue bien défini. Il est défini formellement par **Vincke** comme suit *« une fonction g, définie dans A, prenant ces valeurs dans un ensemble totalement ordonné et représentant les préférences du décideur selon un certain point de vue »* [ALN, 16].

Un critère tout simplement est une expression ou une fonction qualitative ou quantitative qui permet d'exprimer et évaluer les actions.

Pour choisir un critère, il faut respecter les trois conditions suivantes [MAD, 11]:

- **Exhaustivité :** Il ne faut pas oublier un critère.
- **Cohérence :** Il doit y avoir une cohérence entre les préférences locales de chaque critère et les préférences globales. C'est-à-dire que si une action a1 est égale à une action a2 pour tous les critères sauf pour un (ou elle lui est supérieure), ceci signifie que l'action a1 est globalement supérieure à l'action a2.
- **Indépendance** : Il ne doit pas y avoir de redondance entre les critères. Leur nombre doit être tel que la suppression d'un des critères ne permet plus de satisfaire les deux conditions précédentes.

### 4.3. Matrice de performance

Matrice de performance MP ou également matrice d'évaluation ou de jugement désigne un tableau de dimension (m × n) où les colonnes sont les critères, et les lignes sont des vecteurs des performances des actions.

| Critères | Actions (Alternatives) | | | | |
|---|---|---|---|---|---|
| | $a_1$ | $a_2$ | $a_3$ | ………. | $a_n$ |
| $c_1$ | $S_{11}$ | | | | $S_{1n}$ |
| $c_2$ | $S_{21}$ | | | | $S_{2n}$ |
| …… | … | | | | |
| $c_m$ | $S_{m1}$ | | | | $S_{mn}$ |

**Tableau 2** : Structure d'une matrice de performances.

### 4.4. Paramètres subjectifs

Ce sont des paramètres fixés par le décideur, selon l'application et la situation traitée. Ils peuvent être classés en deux catégories : « **paramètres intercritères** » et « **paramètres intracritères** »[HAM, 16].

#### 4.4.1. Paramètres intercritères

Ce sont des paramètres utilisés pour évaluer l'importance relative de chaque critère, on parle souvent de **Poids.** Il s'agit d'un nombre $p_j\{j=1,2,\ldots,m\}$ attribué différemment à chaque critère selon son importance vis-à-vis des autres critères [HAM, 16].

Pour faciliter aux décideurs l'évaluation des critères, l'homme d'étude peut ajouter des méthodes qu'expriment la manière de comparaison et l'évaluation. Parmi ces méthodes, il y a la méthode de l'échelle de Saaty.

❖ **Echelle de Saaty**

Le principe consiste à comparer chaque critère aux autres et introduire un rapport de préférences selon échelle de Saaty (Tableau 3) dans une matrice d'ordre (nc*nc) ou nc : désigne le nombre de critères. Cette matrice est caractérisée par [HAM, 16] :

$\forall i \in [1, nc]$ ; $a_{i,i}=1$ ;

$\forall i,j \in [1, nc^2]$ ; $a_{i,j}=1/a_{j,i}$ avec $i \neq j$ ;

| Valeur | Description |
|---|---|
| 1 | Egal |
| 3 | Un peu plus important |
| 5 | Plus important |
| 7 | Beaucoup plus important |
| 9 | Absolument plus important |

**Tableau 3** : Echelle de Saaty.

#### 4.4.2. Paramètres intracritères

Ils formalisent pour chaque critère l'appréciation subjective de leurs valeurs. On distingue le seuil d'indifférences, le seuil de préférences, et le seuil de véto [HAM, 16].

## 5. Formulation d'un problème décisionnel multicritères

La formulation multicritères d'un problème de décision multicritères est définie par le modèle de (A, A/F, E, Ps ) où [HAM, 16]:

**A** : désigne l'ensemble des actions potentielles envisageables.

**A/F** : est l'ensemble fini des attributs ou critères, généralement conflictuels, à partir desquels les actions seront évaluées [HAM, 16].

**E** : est l'ensemble des évaluations de performances des actions selon chacun des attributs ou critères, c'est-à-dire l'ensemble des vecteurs de performances, un vecteur par action .

**Ps** : désigne l'ensemble des paramètres subjectifs selon le type de la problématique décisionnelle.

## 6. Agrégation multicritères

Il s'agit d'établir un modèle des préférences globales, c'est-à-dire une représentation formalisant de telles préférences relativement à un ensemble A d'actions potentielles, que l'homme d'étude juge approprié au problème d'aide à la décision [HAM, 08]. Il existe trois types d'agrégation multicritères selon B Roy [MAD, 11] :

### 6.1. Agrégation complète

Dans cette approche d'inspiration américaine, les différents critères sont synthétisés dans une seule fonction mathématique monotone (à sens d'évaluation unique). A partir des évaluations des différents critères, la fonction d'optimisation résultante dite d'utilité ou d'agrégation. Produit donc une valeur unique évaluant globalement la solution [OUF, 09].

Peut citer plusieurs méthodes :MAUT (Multiple Attribute Utility Theory), UTA (Utilité Additives), AHP (AnalticHierarchyProcess), etc[OUF, 09].

### 6.2. Agrégation locale

Cette approche repose sur la comparaison des actions deux à deux puis une synthèse des résultats de ces comparaison (c'est d'ailleurs la façon de synthétisé qui diffère entre les méthodes de cette approche). Parmi les méthodes les plus connu d'agrégation partiel, on cite la famille d'ELECTRE (ELimination Et Choix TRaduisantla Réalité)et Prométhée [OUF, 09].

### 6.3. Agrégation partielle

Contrairement aux deux approches précédents où l'on suppose que l'ensemble des actions est fini et de dimension raisonnable, cette approche s'applique à des ensembles d'actions d'une très grande dimension voire infinis lorsque les actions varient en contenu. Partant d'une solution de départ, la technique permet de chercher au voisinage de cette solution s'il n'y a pas de meilleur et ce de manière répétitive [MAD, 11].

Les principales méthodes d'agrégation locale itératives existantes sont : Plm (Programmation Linéaire Multicritère), Stem (Pop), etc. [MAD, 11]

## 7. Conclusion

L'analyse multicritères vise à donner des outils ou des méthodes qui aider à résoudre un problème posé, selon le type de cette problème, et la subjectivité des décideurs.

L'analyse multicritères est une approche très connue, et utilisée pour trouver la solution à des problèmes multicritères avec des points de vue du décideur et les critères souvent contradictoires.